\RequirePackage[l2tabu, orthodox]{nag}
\RequirePackage{snapshot}

\documentclass[10pt,onecolumn]{article}

\makeatletter
\if@twocolumn
  \usepackage[dvips,letterpaper,top=0.5in, bottom=0.5in, left=0.75in, right=0.5in,includefoot,heightrounded]{geometry}
\else
  \usepackage[dvips,letterpaper,margin=1in,includefoot,heightrounded]{geometry}
\fi

\usepackage{srcltx}

\usepackage{amsmath}
\usepackage{amssymb,amsfonts}

\usepackage{abstract}

\usepackage{graphicx}
\usepackage{epsfig}
\usepackage[usenames,dvipsnames]{color}
\usepackage[usenames,dvipsnames]{xcolor}
\usepackage{subfigure}

\usepackage{booktabs}

\usepackage{setspace}
\usepackage{flushend}
\usepackage{multicol}

\usepackage{cite}
\usepackage{url}
\usepackage[normalem]{ulem}

\usepackage{enumerate}

\usepackage{hyperref}

\usepackage[sc,tiny,center]{titlesec}

\usepackage{microtype}



\title{%
Robust Image Watermarking Using Non-Regular Wavelets
}

\author{%
Renato~J.~Cintra%
\thanks{R.~J.~Cintra is with the Department of Statistics, Federal University of Pernambuco, Brazil, e-mail: rjdsc@de.ufpe.br\,.}
\quad and \quad
Todor~V.~Cooklev%
\thanks{%
T.~V.~Cooklev is with the Center for Wireless Communication, Indiana University
Purdue University Fort Wayne, IN, e-mail: tcooklev@ieee.org\,.}
}

\date{}

\newcommand{\myabstract}{%
An approach to watermarking digital images using non-regular wavelets is advanced.
Non-regular transforms spread the energy in the transform domain.
The proposed method leads at the same time to increased image quality and increased
robustness with respect to lossy compression.
The approach provides robust watermarking by suitably creating watermarked messages
that have energy compaction and frequency spreading.
Our experimental results show that
the application of non-regular wavelets,
instead of regular ones,
can furnish a superior
robust watermarking scheme.
The generated watermarked data is
more immune against
non-intentional JPEG and JPEG2000 attacks.
}

\newcommand{\mykeywords}{%
Robust watermarking, non-regular wavelets, JPEG compression
}

\begin{document}

\makeatletter
\if@twocolumn

\twocolumn[%
  \maketitle
  \begin{onecolabstract}
    \myabstract
  \end{onecolabstract}
  \begin{center}
    \small
    \textbf{Keywords}
    \\\medskip
    \mykeywords
  \end{center}
  \bigskip
]
\saythanks

\else

  \maketitle
  \begin{abstract}
    \myabstract
  \end{abstract}
  \begin{center}
    \small
    \textbf{Keywords}
    \\\medskip
    \mykeywords
  \end{center}
  \bigskip
  \onehalfspacing
\fi

\section{Introduction}

Digital watermarking of images has been an active area of research for over a decade.
The main objectives of watermarking are to protect the ownerships of the host images
and to preserve their authenticities.
Watermarking techniques have also been applied
in areas as diverse as
data hiding~\cite{zhichenge2006hiding},
biomedical signal processing~\cite{engin2005ecg,coatrieux2006healthcare},
image indexing~\cite{jiang2002indexing},
image hashing~\cite{cannons2004hash},
and broadcast monitoring~\cite{ming2007am}.
The two main requirements are that watermarks be imperceptible and robust against lossy signal processing.
Watermarking methods that are not robust against
lossy signal processing are called fragile~\cite{lu2003fragile}.
In principle, robust watermarking is performed by inserting a hidden message
--- the watermark ---
into a host image,
such that the hidden message can survive
different types of attacks such as compression, resizing, filtering, re-scanning, or printing.
Detection algorithms are used to retrieve the watermark from watermarked messages~\cite{rey2002survey}.
The robustness of a watermarking algorithm is measured
by the ability to detect successfully the watermark after an attack.

Early watermarking algorithms used spatial domain techniques
which insert watermarked messages into
the spatial domain of the host images~\cite{schyndel1994digital}.
Generally,
inserted messages were pseudorandom sequences similar
to the sequences found in spread spectrum techniques used in
wireless communications systems~\cite{fiebig1993correlation}.
Pseudorandom messages could be obtained by first creating
an 1-D random message with length equal to the number of columns.
Then this random message was circularly shifted with random phase shift for each row.
Mainly,
the watermark could be detected using correlation techniques.
Block-based 2-D signal processing could also be used~\cite{pei2003hybrid}.
Because spread-spectrum pseudo-noise contains high frequency components,
the robustness of spatial techniques are sensitive to compression and lowpass filters.
Nevertheless, it continues to find recent applications~\cite{mukherjee2004spatial}.

Another possible venue is the insertion of the watermark
into the
spectral domain of the subject data.
There are approaches that embed the watermark in the low-frequency subbands
to improve the robustness against a variety of attacks,
e.g. lowpass filtering and compression.
Several transform techniques have been employed to
fulfill this purpose.
Among them, the discrete Fourier transform, the discrete cosine transform,
and a number of wavelet transforms deserve mention.
However,
current investigations in this area
significantly
embrace
the application of wavelets.
Publications~\cite{furht2006techniques,%
potdar2005survey,meerwald2001survey,langelaar2000overview}
describe several wavelet domain watermarking methods and include comparisons.

In this paper, we propose a new watermarking technique that is based
on non-regular wavelet transforms.
The application of non-regular transforms to image processing is novel.
Our experimental results show that non-regular transforms are
superior to regular transforms
for robust watermarking purposes.
The reason that non-regular wavelet filters are
more robust is that they offer energy spreading,
while regular wavelet filters offer energy concentration.

The paper is organized as follows.
Section~\ref{section.nonregular.dwt} presents non-regular discrete wavelet transforms (DWT)
and a watermarking method
using non-regular wavelets is proposed
in Section~\ref{section.using.nonregular}.
In Section~\ref{section.detection},
a statistical method for watermarking detection is elaborated.
Computational experiments and discussions are presented in Section~\ref{section.results}.
Finally, Section~\ref{section.conclusions} summarizes the obtained results.

\section{Non-regular Discrete Wavelet Transforms}
\label{section.nonregular.dwt}

\begin{figure}
\centering
\epsfig{file=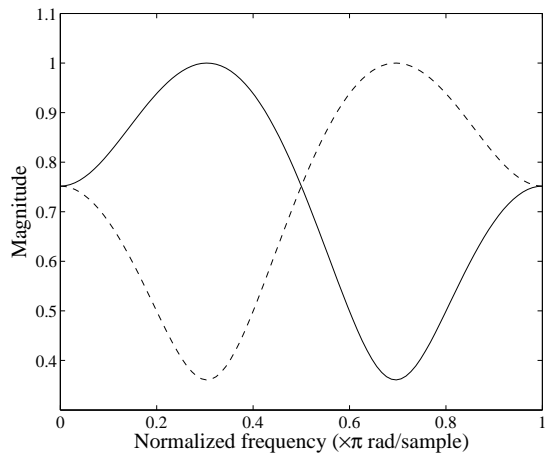, width=0.5\linewidth}
\caption{Magnitude response of the GRS4 filter pair: $H_0$ (solid line) and $H_1$ (dashed line).}
\label{fig1}
\end{figure}

Frequently,
wavelet generation is based on filter bank methods.
Filter banks that perform DWT have always been designed so that
in addition to perfect reconstruction the filters have in some sense good frequency responses,
e.g. in two-channel transforms $H_0(z)$ has always been required to be a good lowpass filter
and $H_1(z)$ --- to be a good highpass filter.
Good lowpass and highpass filters have the energy compaction property.
Provided that $H_0(z)$ is regular,
iterative algorithms for wavelet generation
converge and smooth
functions can be obtained~\cite{strang1996wavelets}.
If $H_0(z)$ is non-regular,
then
iterative processes fail to converge~\cite{daubechies1992lectures}.
For example, complementary or Golay-Rudin-Shapiro (GRS) polynomials,
which have coefficients equal to $+1$ or $-1$~\cite{golay1961series},
can be utilized to design
non-regular wavelet filters.
Two polynomials, $H_{00}(z)$ and $H_{01}(z)$, each with $l$ coefficients, are complementary if
\begin{align}
\label{eq1}
H_{00}(z)H_{00}(z^{-1})+H_{01}(z)H_{01}(z^{-1})=2l.
\end{align}
Then the filter pair
$H_0(z) = H_{00}(z^2) + z^{-1}H_{01}(z^2)$,
and
$H_1(z) = - z^{-N}H_0(-z^{-1})$
defines an orthogonal wavelet transform.
In particular, the non-regular filters
$h_0= \begin{bmatrix}1 &1 &1 &-1\end{bmatrix}$
and
$h_1 = \begin{bmatrix}-1 &-1& 1 &-1\end{bmatrix}$,
called here GRS4 filters, have frequency responses as shown in Figure~\ref{fig1}.
For instance, this pair of filters induce the generation of the GRS4 wavelets.
It is worth to note that these filters do not offer energy compaction, but offer energy spreading.
Starting from a kernel with four coefficients,
other wavelet filters can be obtained recursively.
These non-regular filters have been used in other communications applications~\cite{wornell1996emerging},
where energy spreading is desirable.

\begin{figure}
\centering
\subfigure[]{\epsfig{file=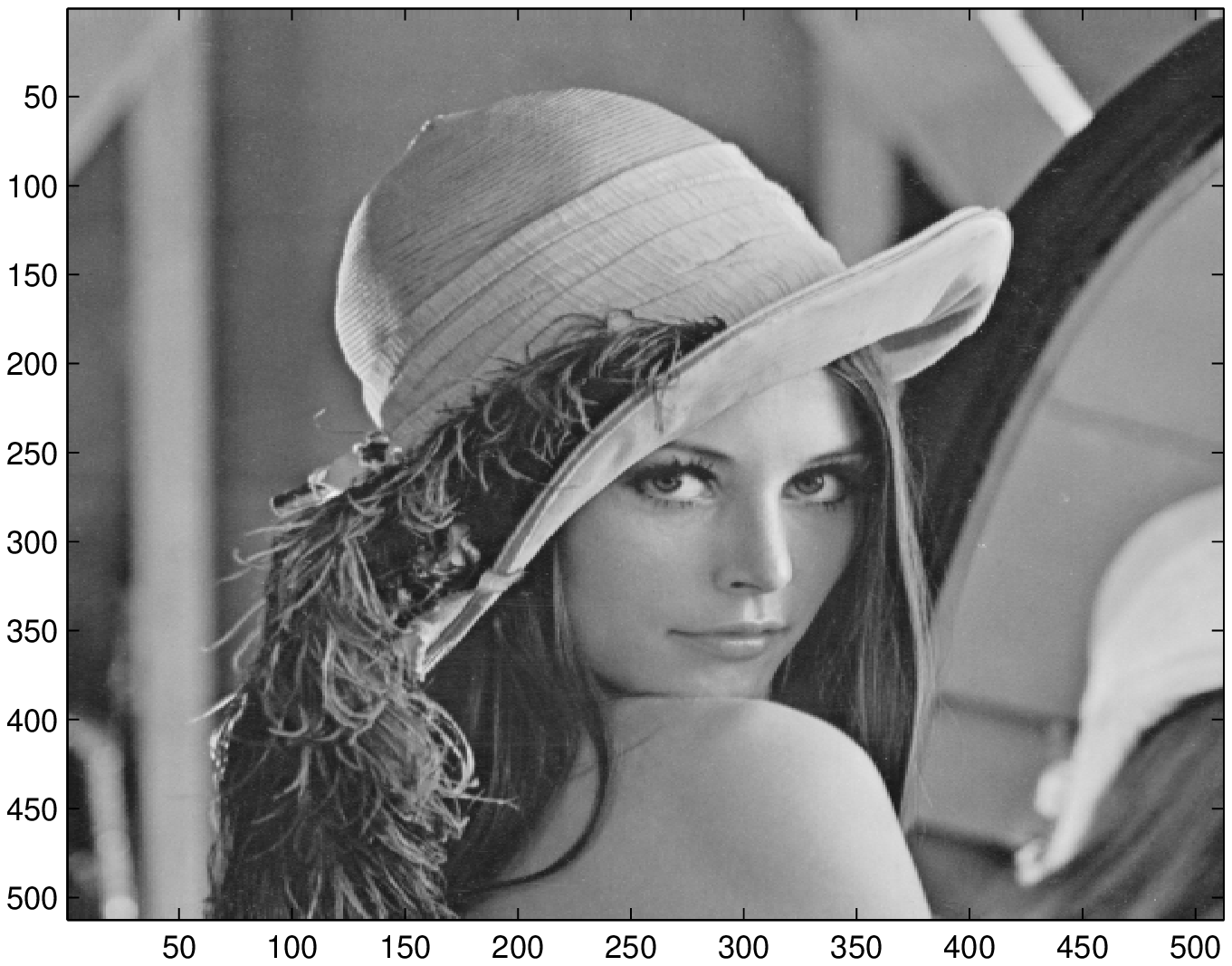, width=0.48\linewidth}}
\subfigure[]{\epsfig{file=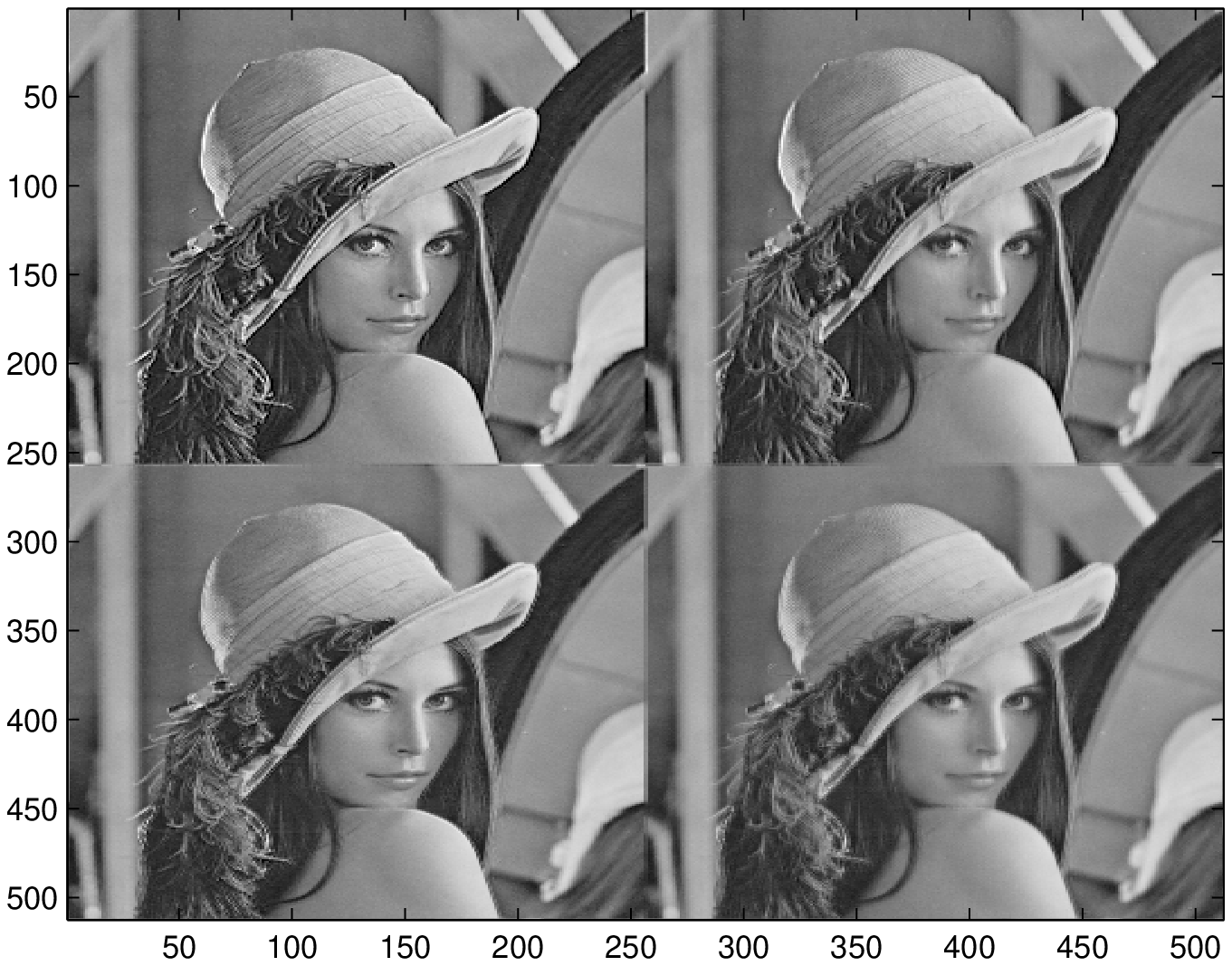, width=0.48\linewidth}} \\
\subfigure[]{\epsfig{file=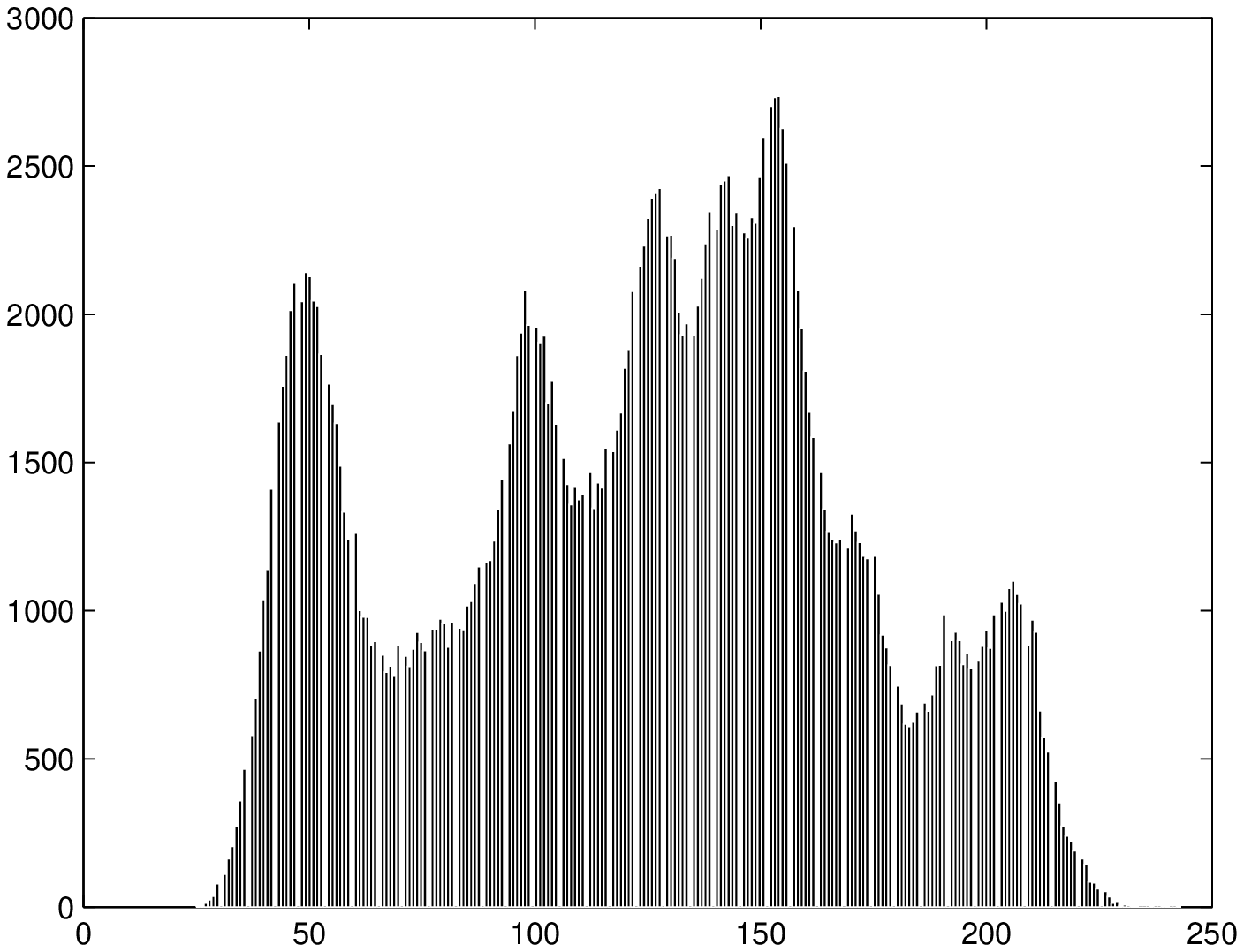, width=0.42\linewidth}}
\subfigure[]{%
\begin{tabular}[b]{cc}
\epsfig{file=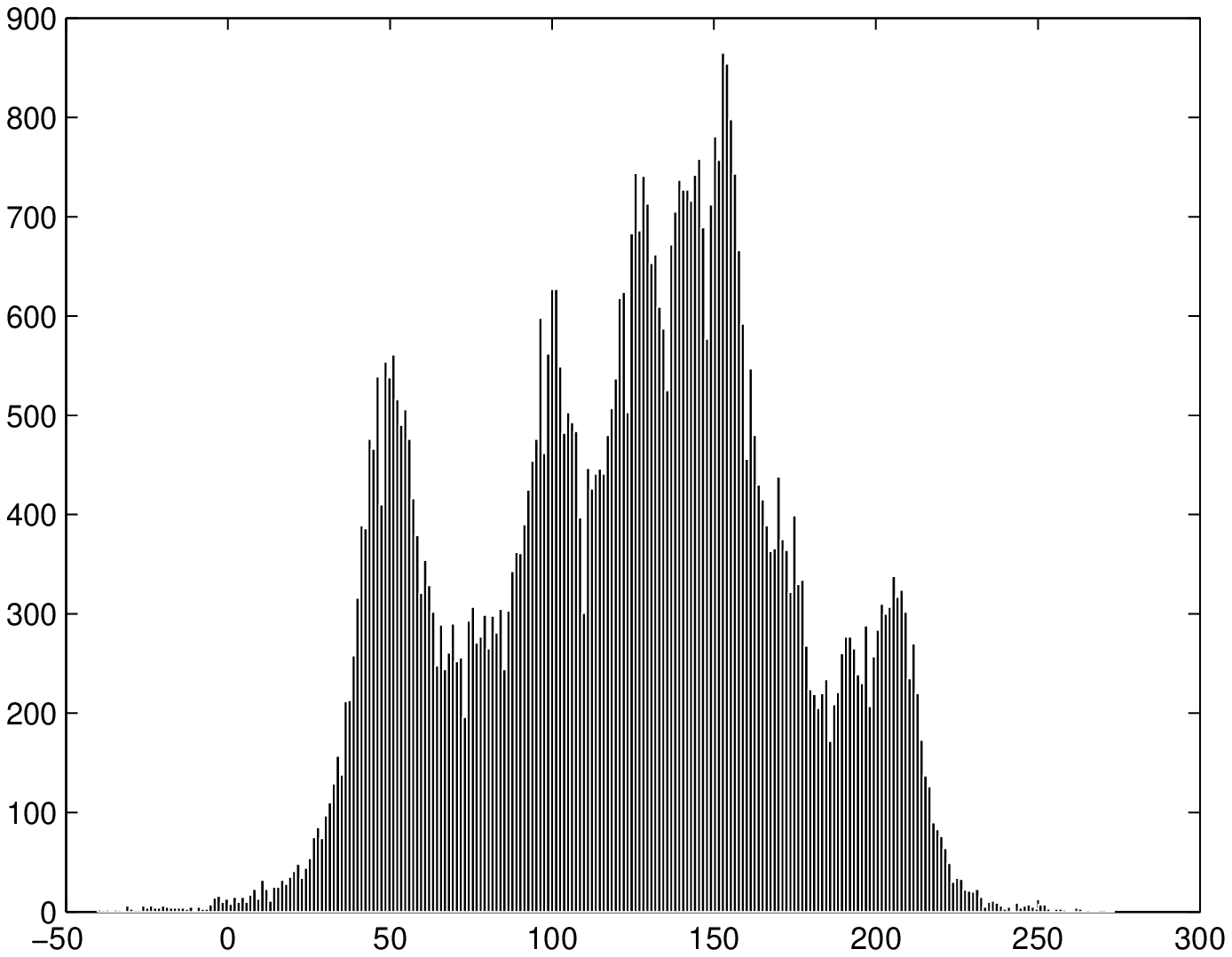, width=0.23\linewidth}
&
\epsfig{file=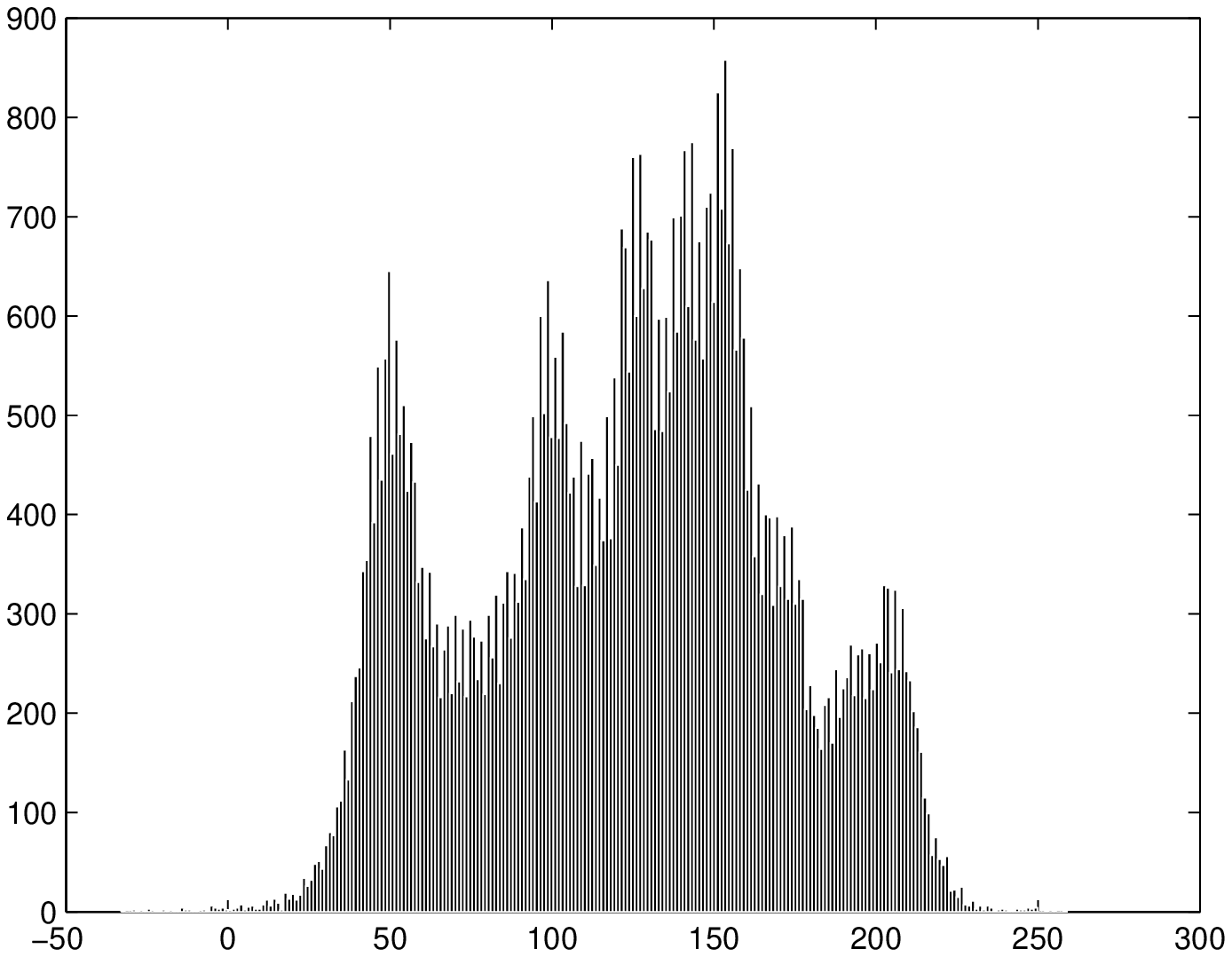, width=0.23\linewidth}
\\
\epsfig{file=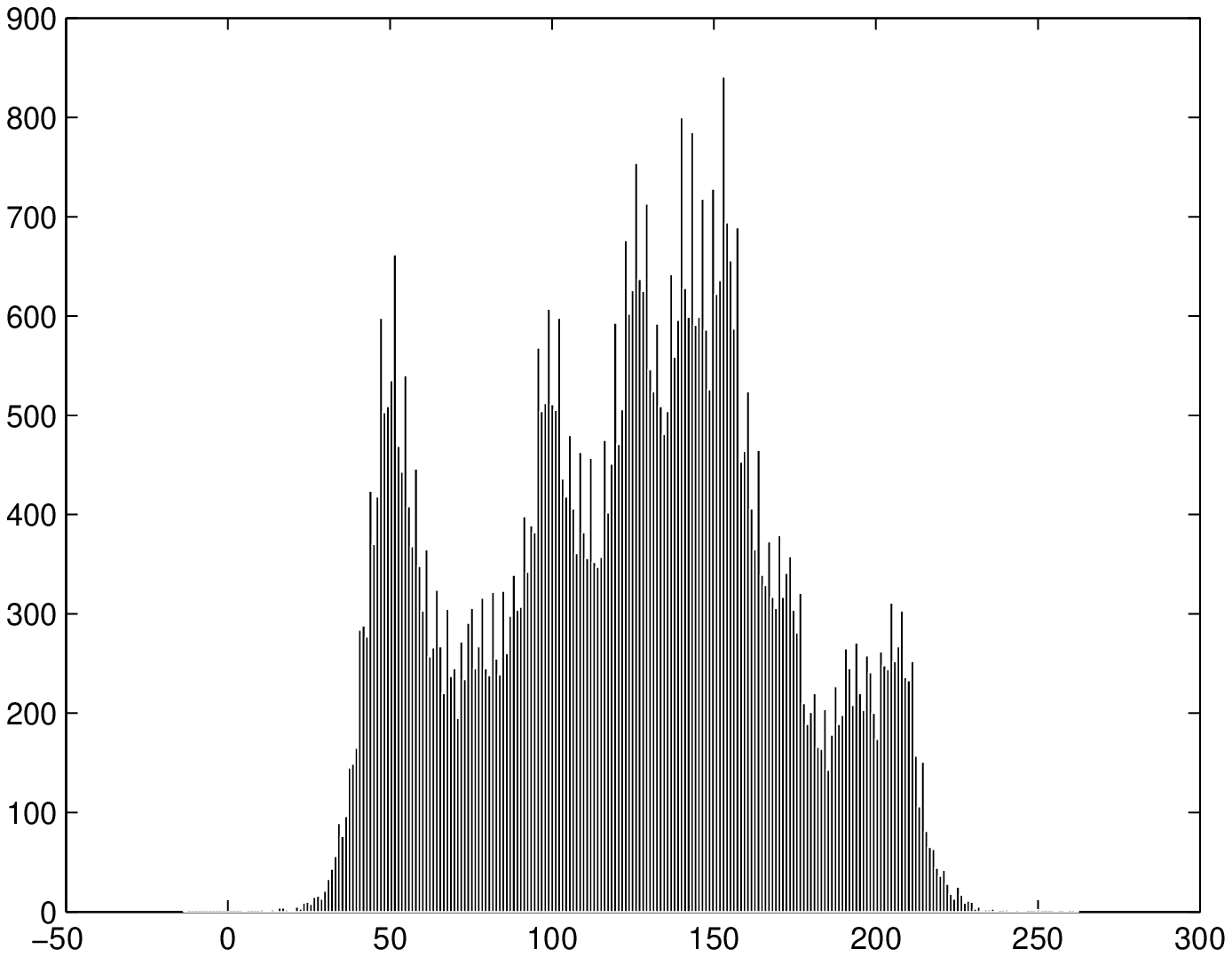, width=0.23\linewidth}
&
\epsfig{file=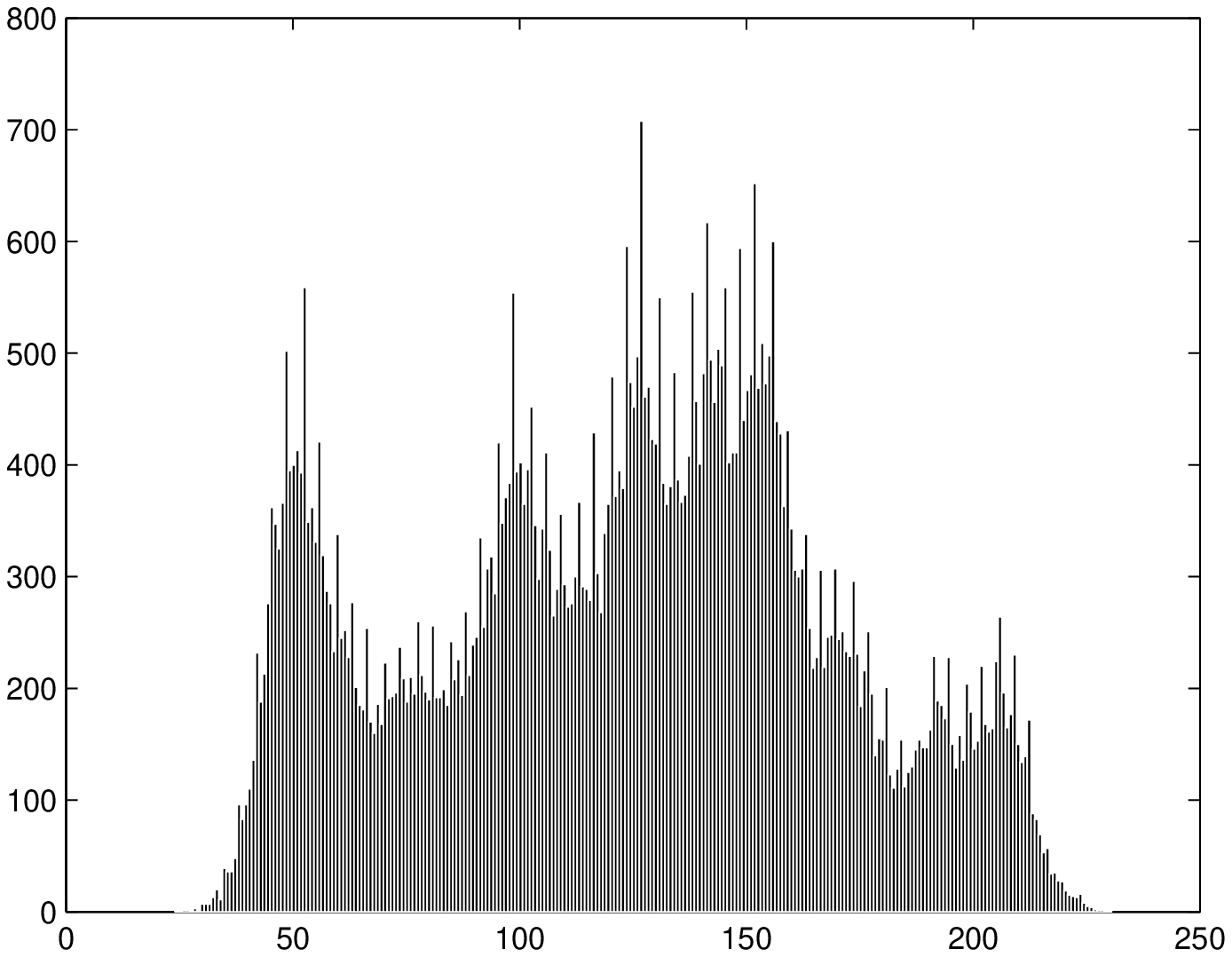, width=0.23\linewidth}
\end{tabular}%
}
\caption{(a) Original image, (b) subbands of the decomposed image,
(c) histogram of the original image, and (d) histogram of each subband, respectively.}
\label{fig2}
\end{figure}

Unlike regular filters, when non-regular filters are used,
a given signal and its DWT share similar statistical properties.
To illustrate this, Figure~\ref{fig2}(a) shows the Lena image in the spatial domain,
as well as in the transform domain furnished by the GRS4 wavelet (Figure~\ref{fig2}(b)).
Clearly, the non-regular transform spreads the energy in all subbands.
In Figure~\ref{fig2}(c) and (d), histograms of the original image and the GRS4 DWT subbands
are depicted, exhibiting noticeable visual similarity.
In order to quantify this behavior,
techniques for measuring the statistical proximity
between two
probability distributions were considered.
A useful tool for this purpose
is the Jensen-Shannon divergence~\cite{lin1991measures}.
This particular divergence is based on
the Kullback-Leibler divergence~\cite{eguchi2006interpreting}.

Extensive computer simulations
have been performed
with 20 different test images
from the USC-SIPI image database~\cite{USCSIPI}.
Then the Jensen-Shannon divergence among each image
and its GRS4 DWT subbands was calculated.
Table~\ref{tab1} shows the average of the obtained
values for the Jensen-Shannon divergence.
Note that the distances are extremely small,
clearly indicating a high statistical proximity among
the examined images and their GRS4 DWT subbands.
Other GRS wavelets also offer comparable results.

The energy spreading property of the non-regular filter banks
explains the obtained statistical similarities.
In fact,
this information theoretical proximity between
spatial and transform domains
is explored to
propose a new robust watermarking scheme.

\begin{table}
\centering
\caption{The average Jensen-Shannon divergence (JSD) among selected
original images and their GRS4 DWT subbands, respectively}
\label{tab1}
\begin{tabular}{c|ccccc}
\hline
JSD ($\times 10^{-2}$)   &    Original & LL    & LH     & HL     & HH \\
\hline
Original &         0   & -       & -        & -        & -  \\
LL       &    3.2014   &      0  & -        & -        & -  \\
LH       &    2.5290   & 0.7972  &       0  & -        & -  \\
HL       &    2.3966   & 0.9731  &  0.2843  &       0  & -  \\
HH       &    2.6530   & 2.5759  &  0.8467  &  0.6518  &       0 \\
\hline
\end{tabular}
\end{table}

\section{Image Watermarking Using Non-regular DWT}
\label{section.using.nonregular}

\begin{figure}
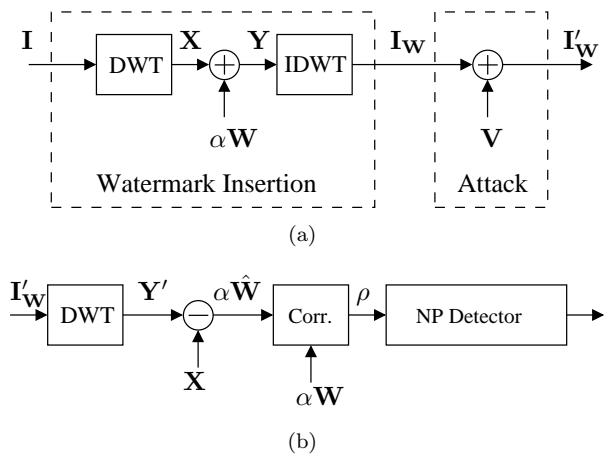

\centering
\subfigure[]{\input{wm_insert_attack.pstex_t}}\\
\subfigure[]{\input{wm_detector.pstex_t}}
\caption{Block diagram of the proposed watermarking scheme: (a) watermark insertion and attack,
(b) watermark detection.}
\label{fig3}
\end{figure}

The block diagram of the proposed
watermarking algorithm is shown in Figure~\ref{fig3}.
The scheme includes three generic steps:
watermark insertion, attack, and detection.
The watermark insertion begins by computing the DWT of the input image
to obtain the low-low (LL), low-high (LH), high-low (HL), and high-high (HH) subbands.
The watermark message is inserted in the wavelet domain by
\begin{align}
\label{eq2}
\mathbf{Y} = \mathbf{X} + \alpha \mathbf{W},
\end{align}
where $\mathbf{Y}$ is the watermarked data in transform domain,
$\mathbf{X}$ is the wavelet transform of the original image~$\mathbf{I}$,
$\mathbf{W}$ is the watermark message,
and $\alpha$ is a scalar value associated to the watermark strength.
Usually,
the watermark can be embedded in all subbands except LL,
or --- as discussed here --- when non-regular filters are used,
in all subbands, including LL.
The size of the watermarking message is equal
to the size of the image when the watermark is embedded in all subbands.
When the watermark is inserted in the LH, HL, and HH subbands,
the size of the watermarking message is equal to the combined size of these subbands.
The watermark strength $\alpha$ should be chosen to satisfy
the two requirements of image quality and robustness.

\begin{figure}
\centering
\epsfig{file=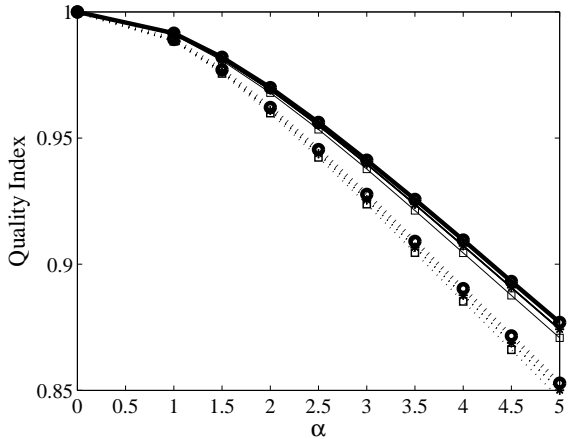, width=0.5\linewidth}
\caption{Quality Index vs. watermark strength for different wavelets:
Daubechies-4 ($\times$), Daubechies-8 ($+$), Coiflet-6 ($\ast$), biorthogonal 6/2 ($\Box$),
and GRS4 ($\circ$).
Watermark was inserted either in all subbands (dotted lines),
or only in the high subbands (solid lines).}
\label{fig4}
\end{figure}

The first requirement is that the watermarked image be indistinguishable from the original image.
While this can be verified subjectively using just-noticeable-difference~\cite{jayant1993models},
the quality of the watermarked image can be calculated using the universal quality index
(UQI) proposed in~\cite{wang2002universal}.
Although the UQI is not based on any HVS model,
it outperforms, for example the mean squared error and the peak signal-to-noise ratio, as a measure of quality.
It is accepted that a UQI value of $0.9$ or greater generally indicates
an imperceptibly distorted image.
Based on a set of twelve standard images available
at the {USC-SIPI} image database~\cite{USCSIPI},
the average UQI of the images as a function of
the watermark strength $\alpha$,
for several wavelets with and without using the subband LL,
was calculated and shown in Figure~\ref{fig4}.
Therefore, a reliable range for the selection of the watermark strength is $\alpha \leq 3.5$.
Judiciously, $\alpha$ was set to three in all instances of this work.
Figure~\ref{fig4} also clearly indicates that non-regular filters lead
to higher values of the universal quality index.
Additionally,
when regular filters are used and the watermark is inserted in the LL subband,
the image quality is lower, as the watermark becomes noticeable.
Posterior to the insertion of the watermark
according to Equation~\ref{eq2},
the computation of
the inverse DWT of $\mathbf{Y}$
furnished
the watermarked image $\mathbf{I_W}$.

In the next step,
the watermarked image
is submitted to non-intentional attacks,
modeled by lossy signal processing.
Specifically,
both JPEG and JPEG2000 compression schemes are employed.
Unlike JPEG coding, JPEG2000 is a wavelet based scheme.
The compression attack can be recognized as an additive
noise signal $\mathbf{V}$, which produces a corrupted
image $\mathbf{I_W'} = \mathbf{I_W} + \mathbf{V}$.

The final step consists of a statistical analysis aiming the estimation and detection
of the watermark message.
The estimated watermark $\hat{\mathbf{W}}$ is obtained using a non-blind approach,
in which the original image is required.
After a forward DWT application on $\mathbf{I_W'}$,
the quantity $\mathbf{Y'}$ is returned and
the watermark message is estimated by the following expression:
\begin{align}
\alpha \hat{\mathbf{W}} = \mathbf{Y'} - \mathbf{X}.
\end{align}

Once the watermark message is estimated, it is necessary to
establish whether it matches the
authentic watermark $\mathbf{W}$.
The detection phase is based on
the correlation coefficient $\rho$
between $\mathbf{W}$ and $\hat{\mathbf{W}}$.
This quantity is evaluated as follows
\begin{align}
\label{eq3}
\rho
=
\frac{\langle \mathbf{W}, \hat{\mathbf{W}} \rangle}%
{\| \mathbf{W} \|_2  \| \hat{\mathbf{W}} \|_2},
\end{align}
where $\langle \cdot, \cdot \rangle$ denotes the usual inner product and
$\| \cdot \|_2$ is the $L^2$ norm.

In this study,
the watermarks were generated by random binary sequences of $+1$ and $-1$
with equal probabilities.
A same random message is added to each subband.
Observe that, for a fixed image,
each step of the watermarking procedure is deterministic.
Then,
a suggested approach to investigate the behavior
of $\rho$
is to perform the watermarking procedure
employing different random watermarks.
Therefore,
the resulting $\rho$
becomes a random variable to be analyzed.

\section{Detection Method}
\label{section.detection}

According to an observed value of $\rho$,
a binary hypothesis test was taken in consideration
attempting to decide between the following
two hypothesis:
(i) the null hypothesis ($\mathcal{H}_0$), corresponding to the absence of
a correct watermark,
and
(ii) the alternative hypothesis ($\mathcal{H}_1$),
implying the
existence of a matching watermark.
It is worth to note that the
complete absence of the watermark, i.e. $\mathbf{W}=\mathbf{0}$,
is categorized as a mismatch.

Using a conservative approach,
a detector based on
the Neyman-Pearson (NP) lemma was devised~\cite{kay1998detection}.
The NP lemma establishes a threshold based test
that maximizes the probability of detection $P_D$
subject to a prescribed probability of false alarm $P_{FA}$.
In symbols,
for a given correlation coefficient $\rho_0$,
the decision rule is to decide $\mathcal{H}_1$
if
\begin{align}
L(\rho_0)
\triangleq
\frac{p(\rho_0;\mathcal{H}_1)}{p(\rho_0; \mathcal{H}_0)}
>
\gamma,
\end{align}
otherwise, decide $\mathcal{H}_0$.
The function $L(\cdot)$ is the likelihood ratio,
$p(\cdot;\mathcal{H}_0)$ and $p(\cdot; \mathcal{H}_1)$
are the probability density functions (pdfs) of $\rho$ for each hypothesis, respectively,
and the threshold $\gamma$ is given by~\cite{kay1998detection}
\begin{align}
\label{gamma}
\int_{\{x:L(x)>\gamma\}} p(x; \mathcal{H}_0) \mathrm{d}x = P_{FA}.
\end{align}

Unfortunately, the NP detector assumes that
the pdfs of $\rho$, for each hypothesis, are known.
However,
no analytical expressions for the above mentioned pdfs are available.
Consequently,
exhaustive numerical simulations were performed to
obtain the empirical pdfs of $\rho$,
$p'(\rho;\mathcal{H}_0)$ and $p'(\rho; \mathcal{H}_1)$,
as approximations for
$p(\rho;\mathcal{H}_0)$ and $p(\rho; \mathcal{H}_1)$,
respectively~\cite{duda2001pattern}.

\begin{figure}
\centering
\subfigure[]{\epsfig{file=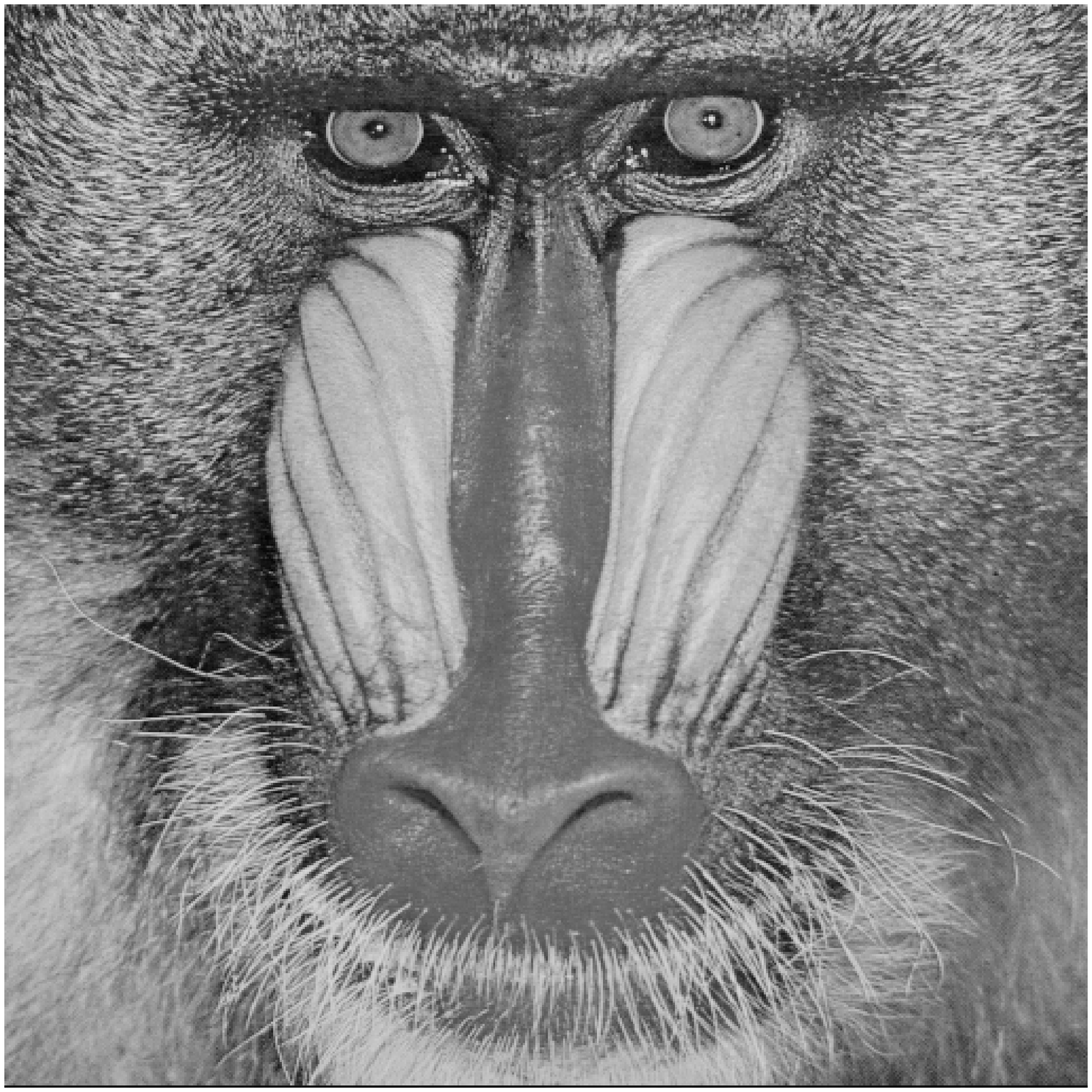, width=0.175\linewidth}}
\subfigure[]{\epsfig{file=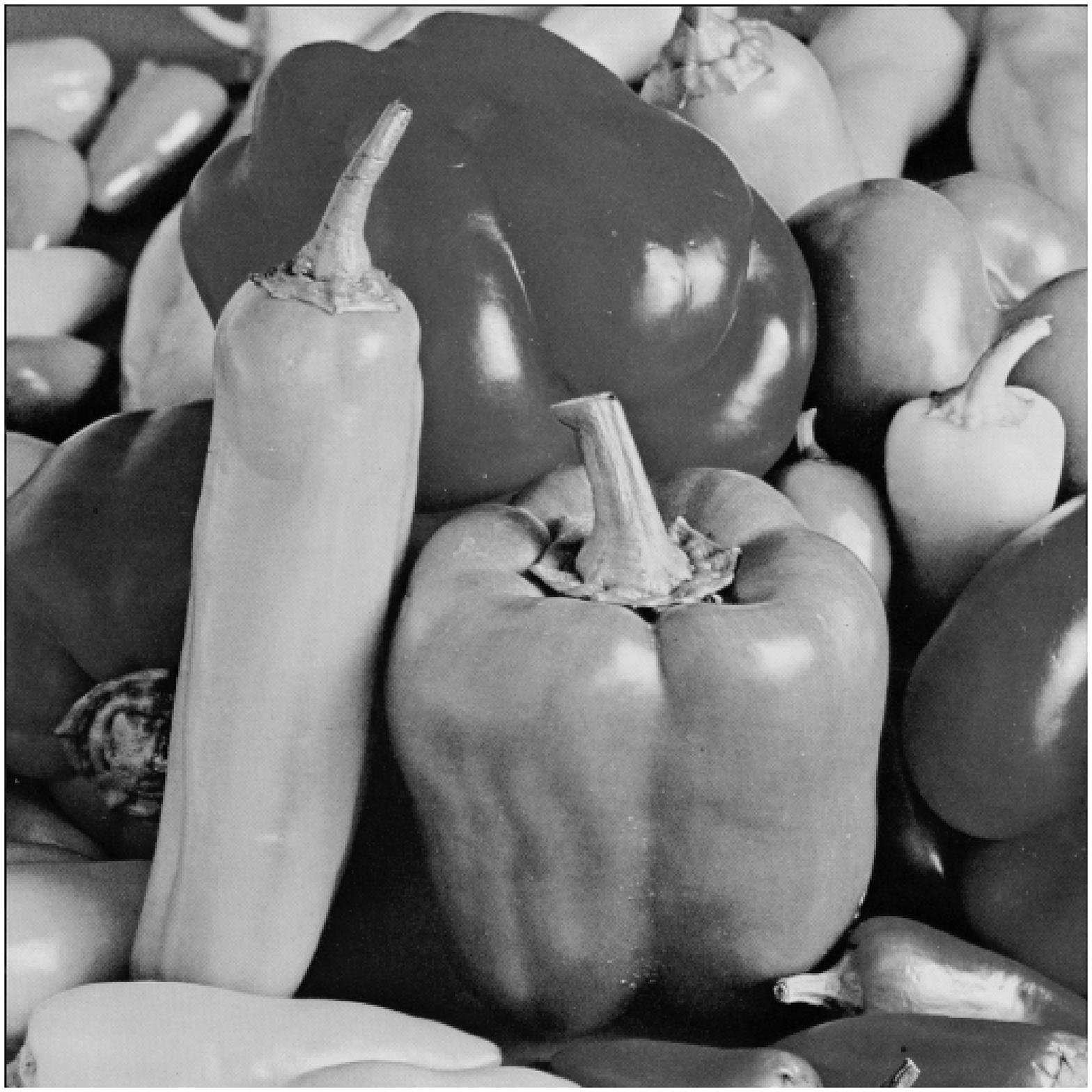, width=0.175\linewidth}}
\subfigure[]{\epsfig{file=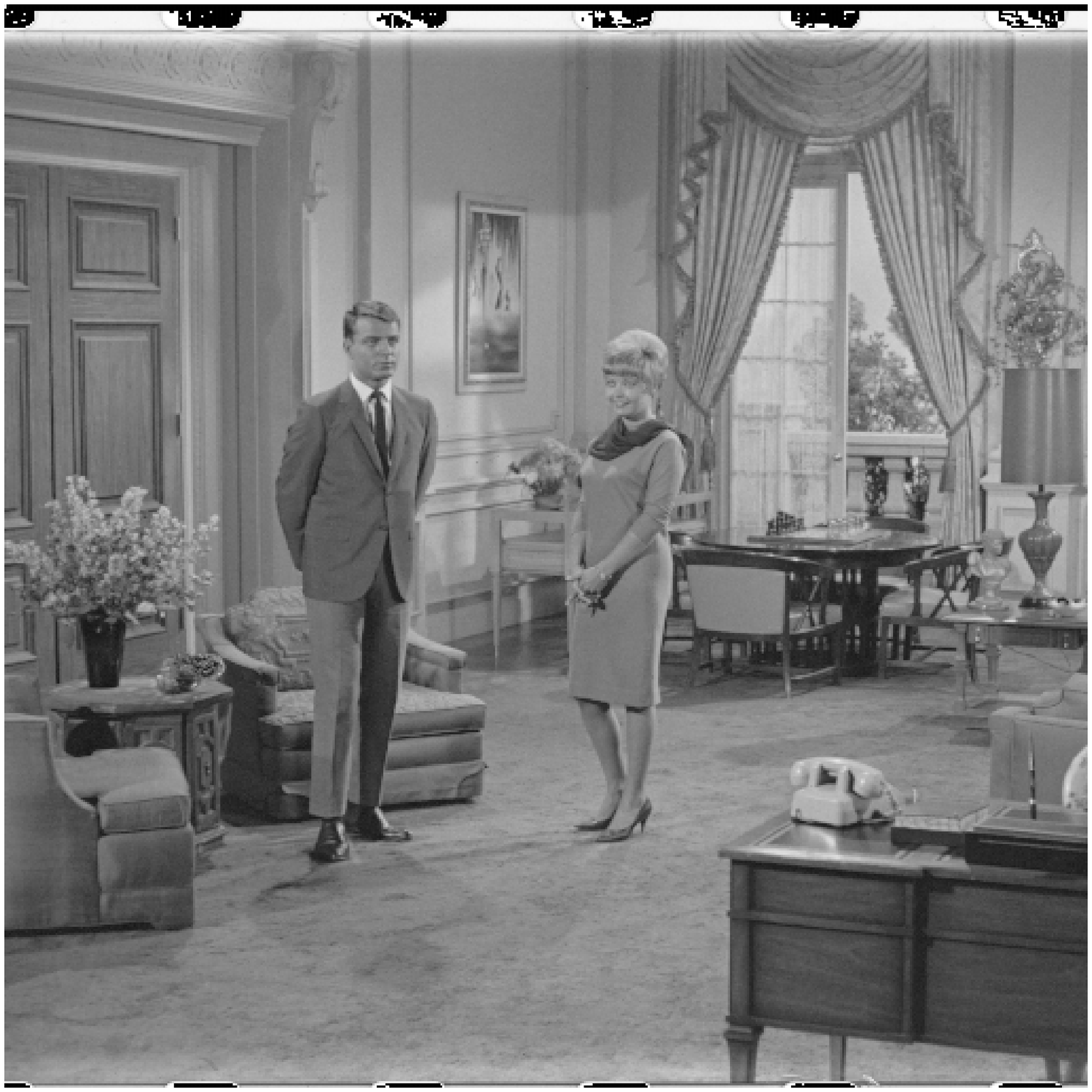, width=0.175\linewidth}}
\subfigure[]{\epsfig{file=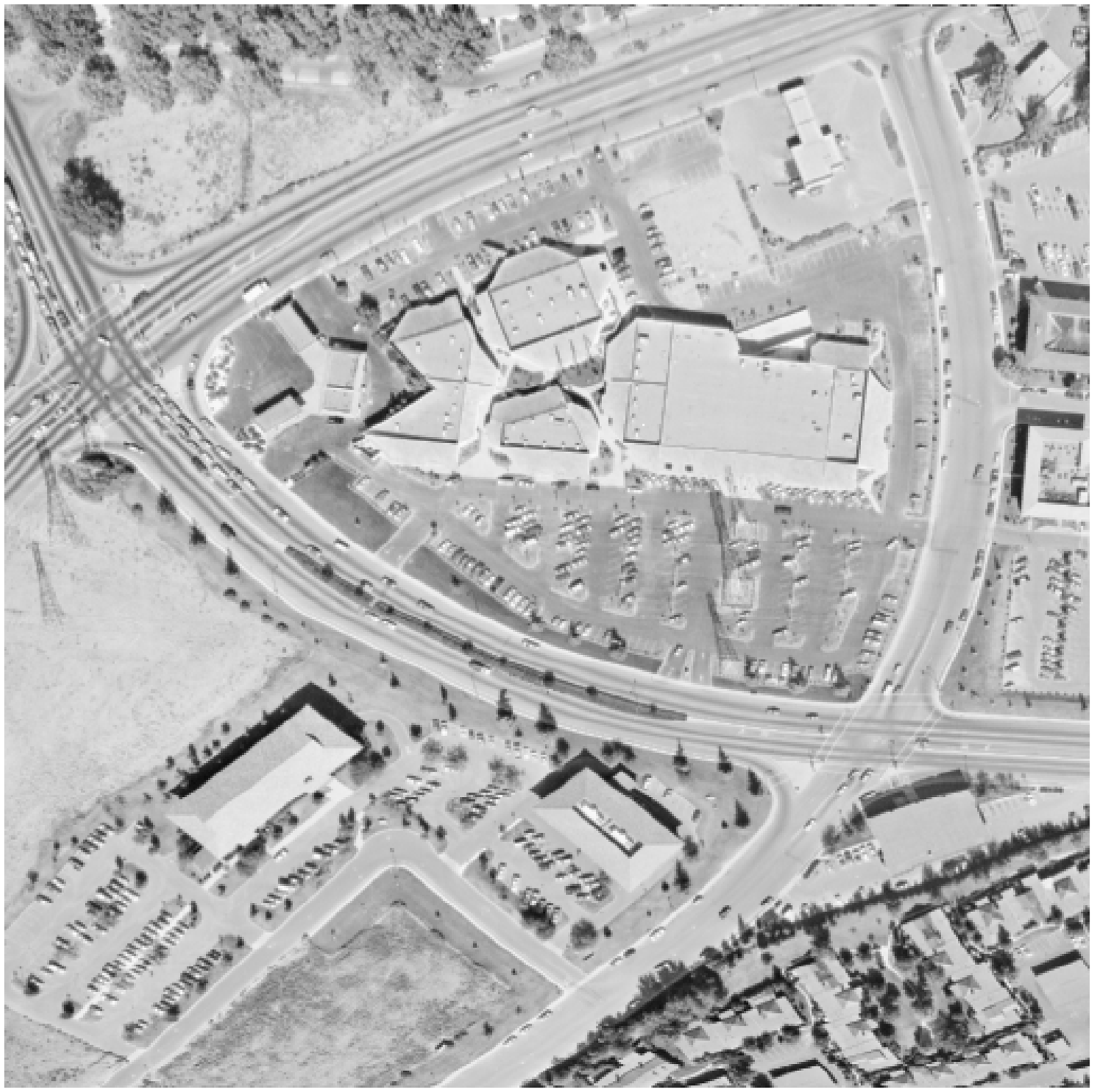, width=0.175\linewidth}} \\
\subfigure[]{\epsfig{file=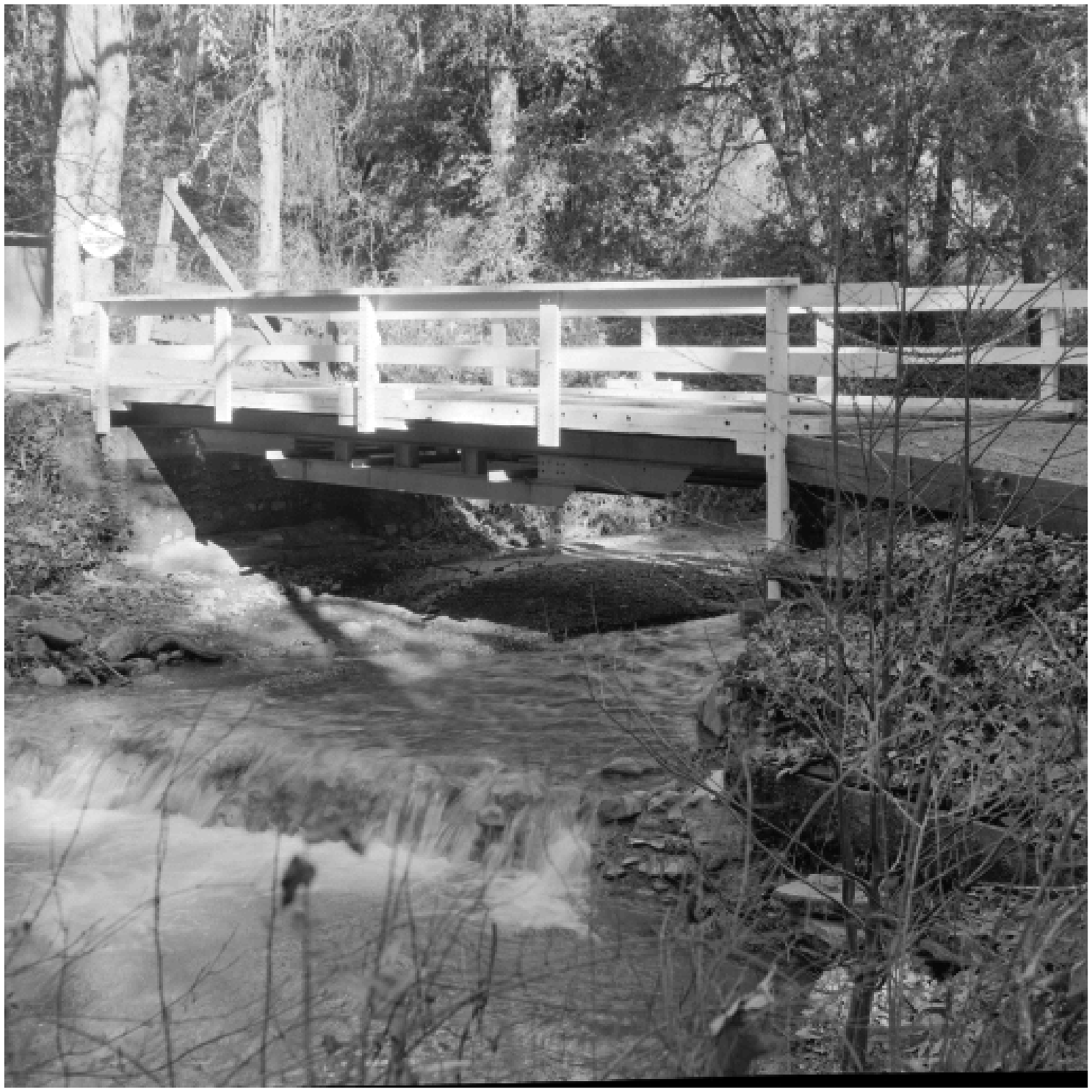, width=0.175\linewidth}}
\subfigure[]{\epsfig{file=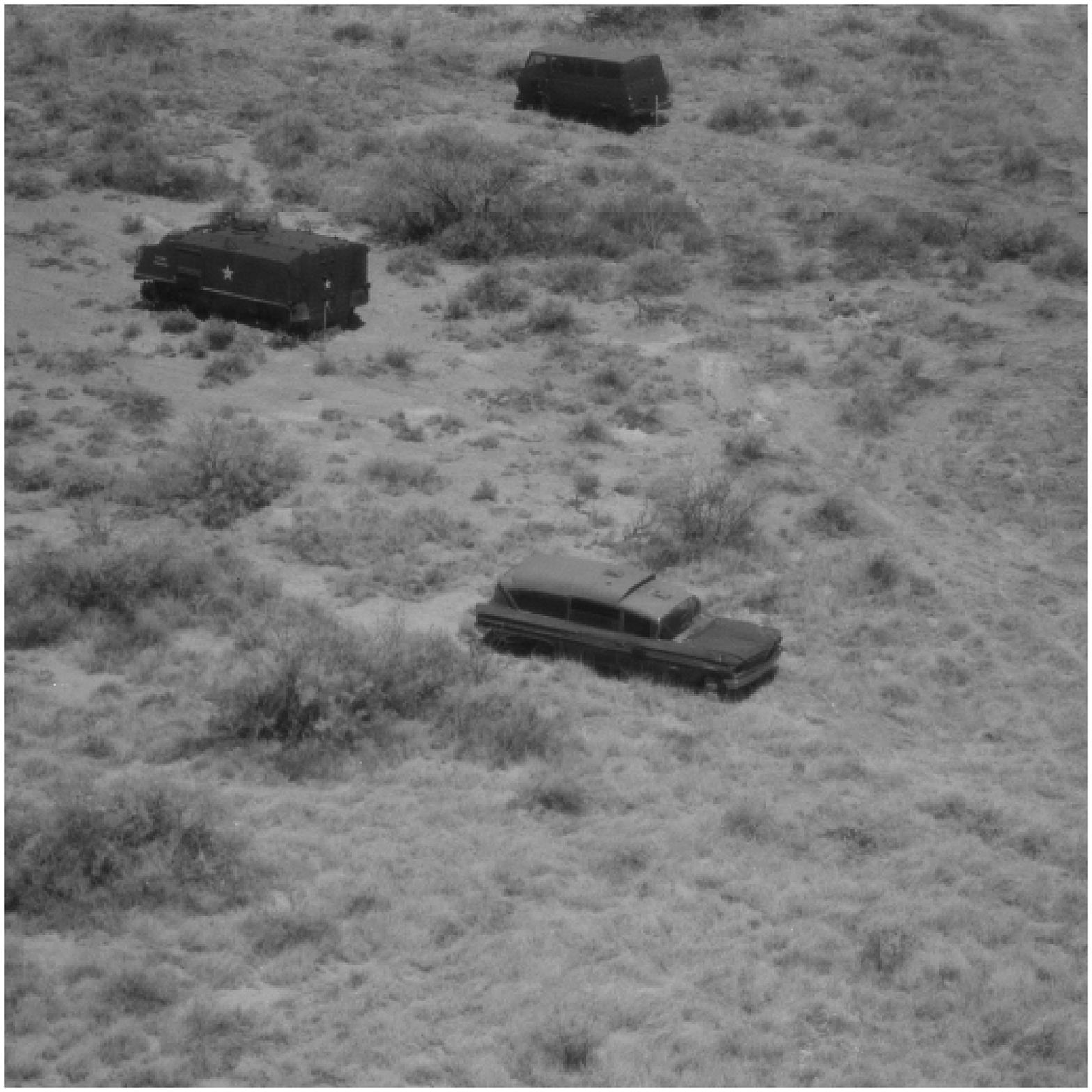, width=0.175\linewidth}}
\subfigure[]{\epsfig{file=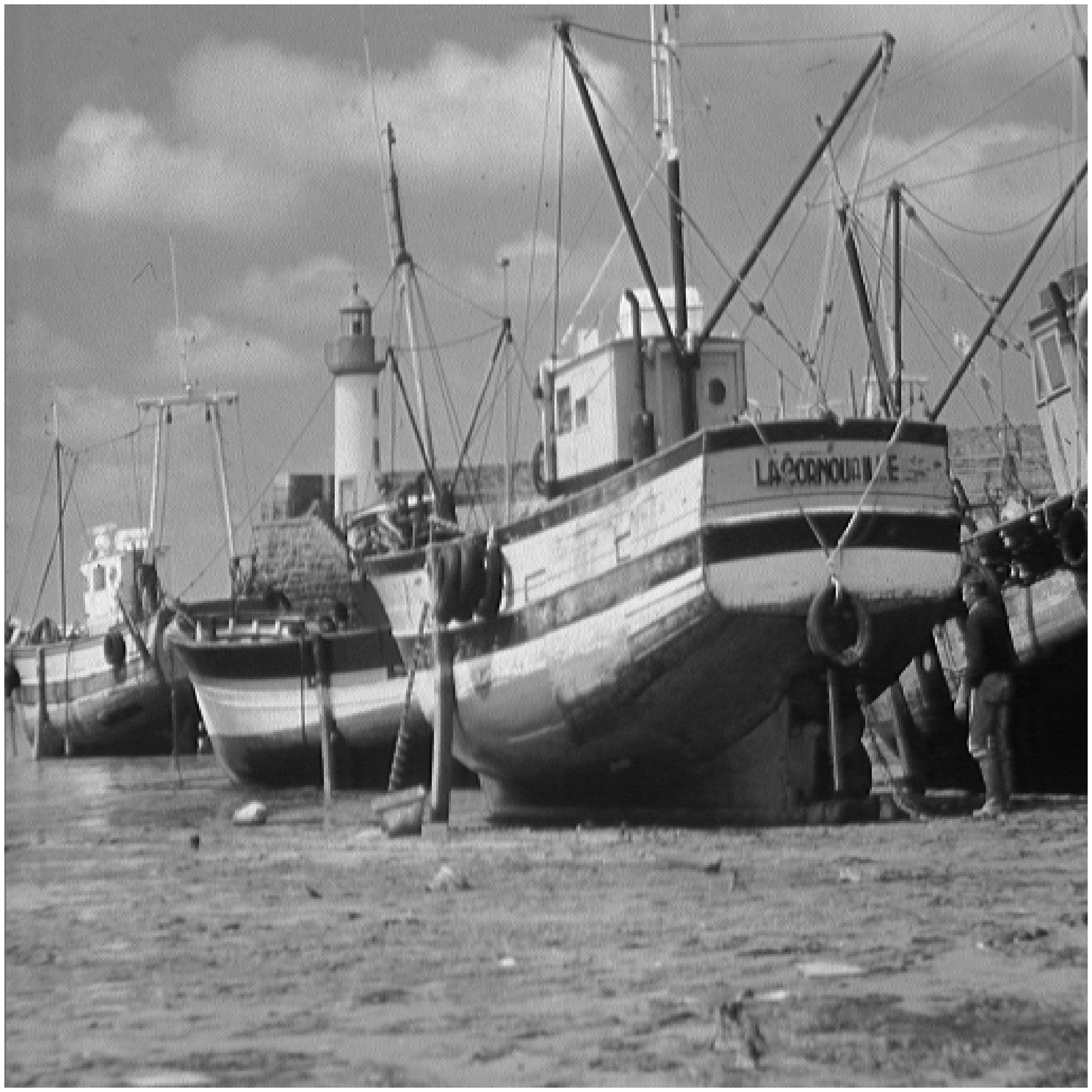, width=0.175\linewidth}}
\subfigure[]{\epsfig{file=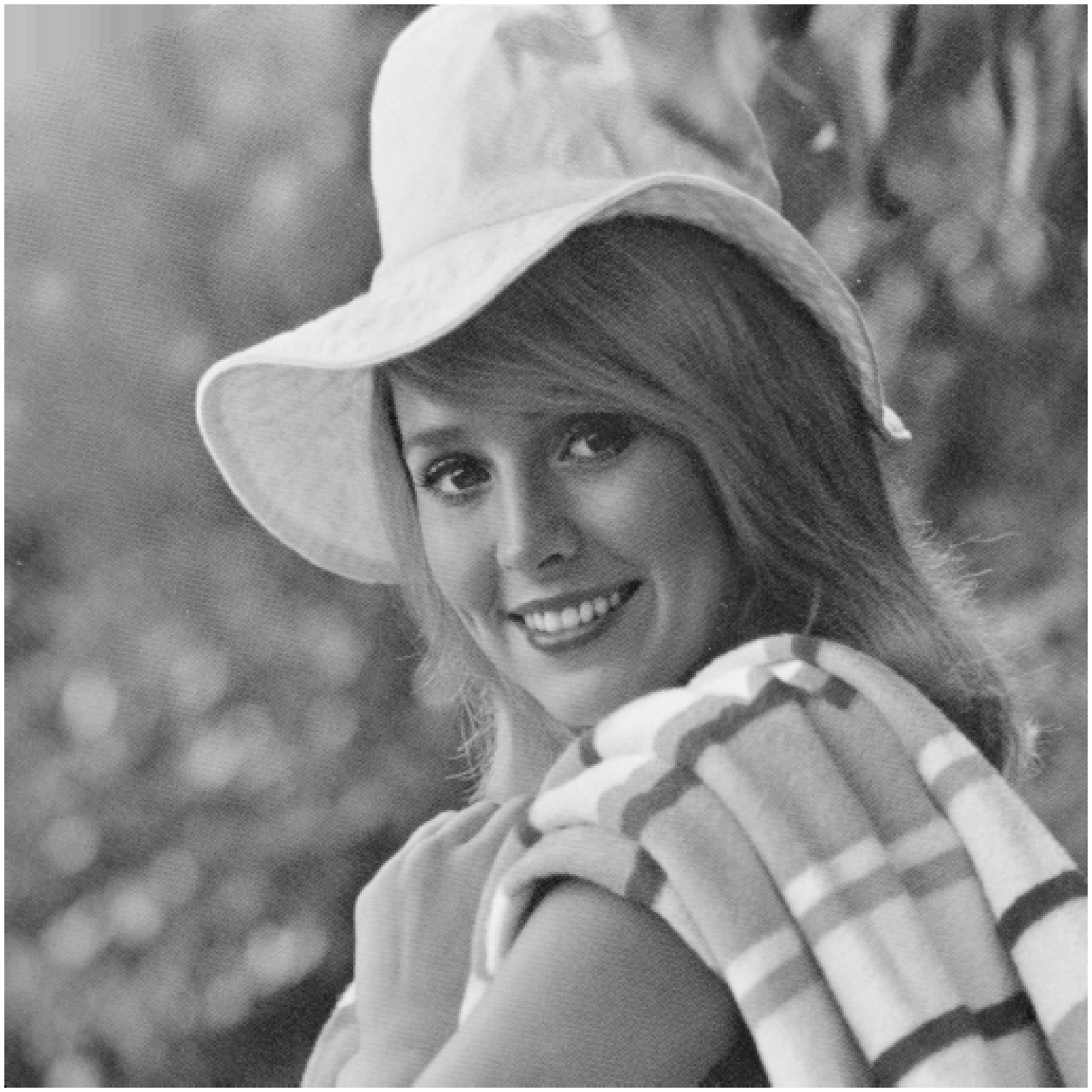, width=0.175\linewidth}} \\
\subfigure[]{\epsfig{file=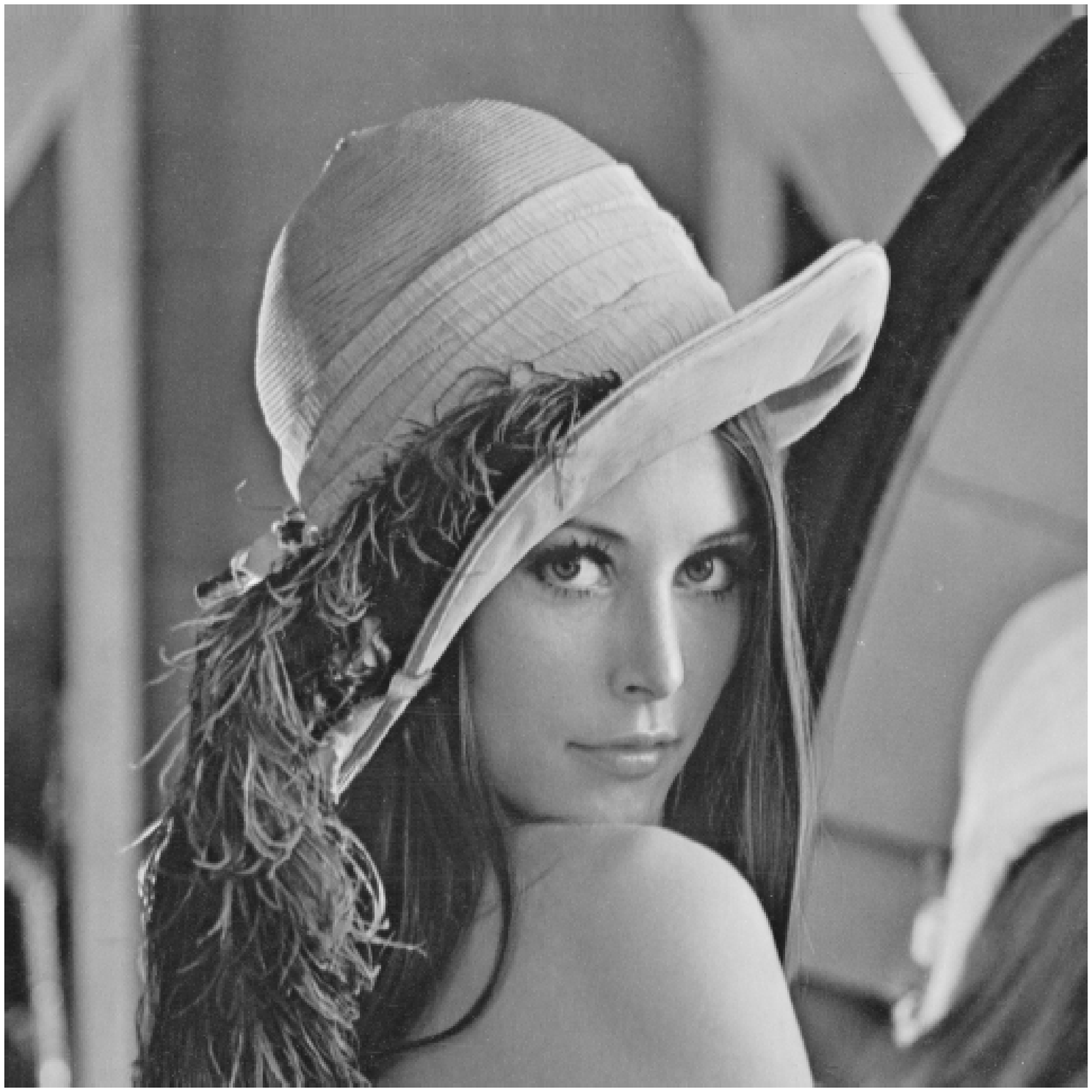, width=0.175\linewidth}}
\subfigure[]{\epsfig{file=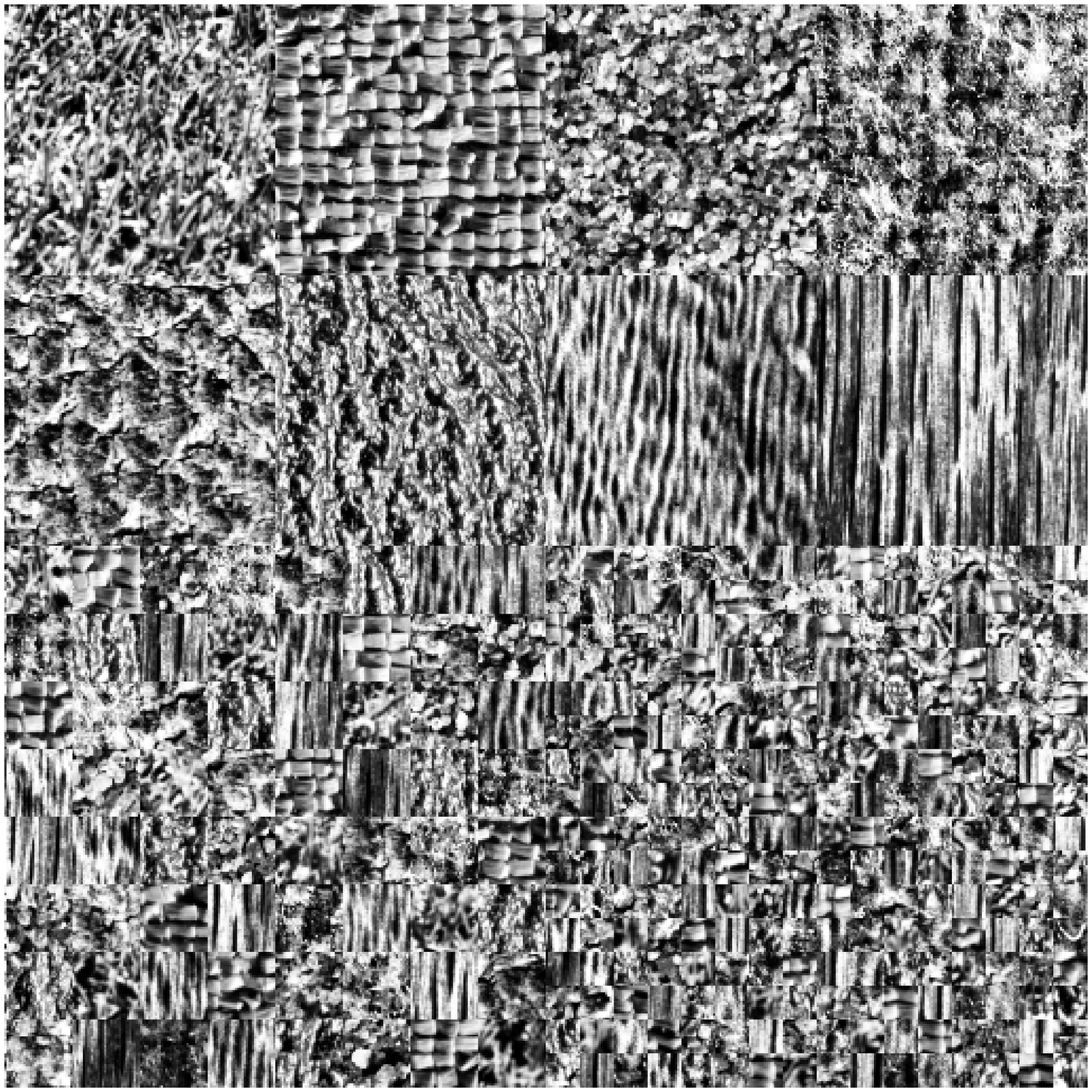, width=0.175\linewidth}}
\subfigure[]{\epsfig{file=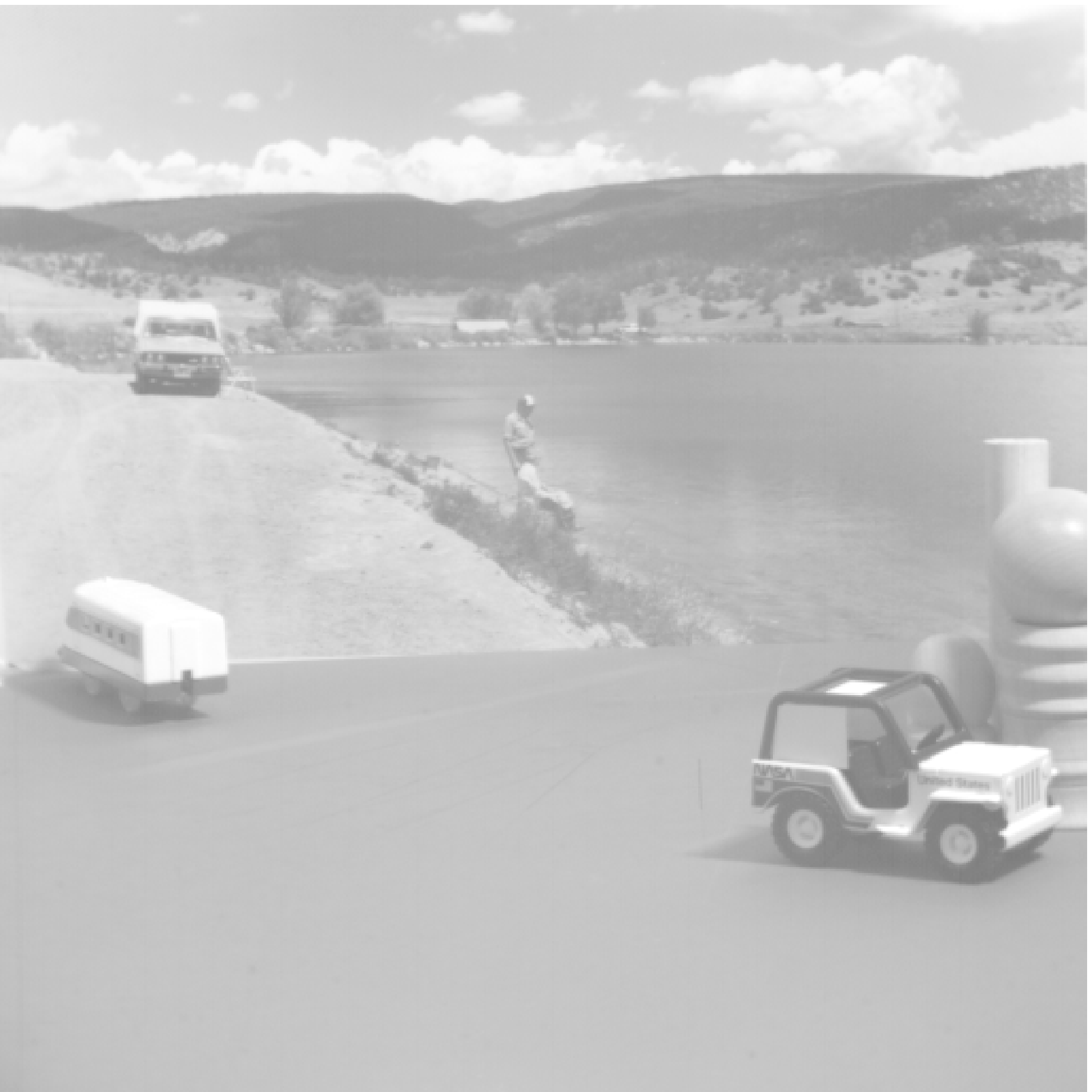, width=0.175\linewidth}}
\subfigure[]{\epsfig{file=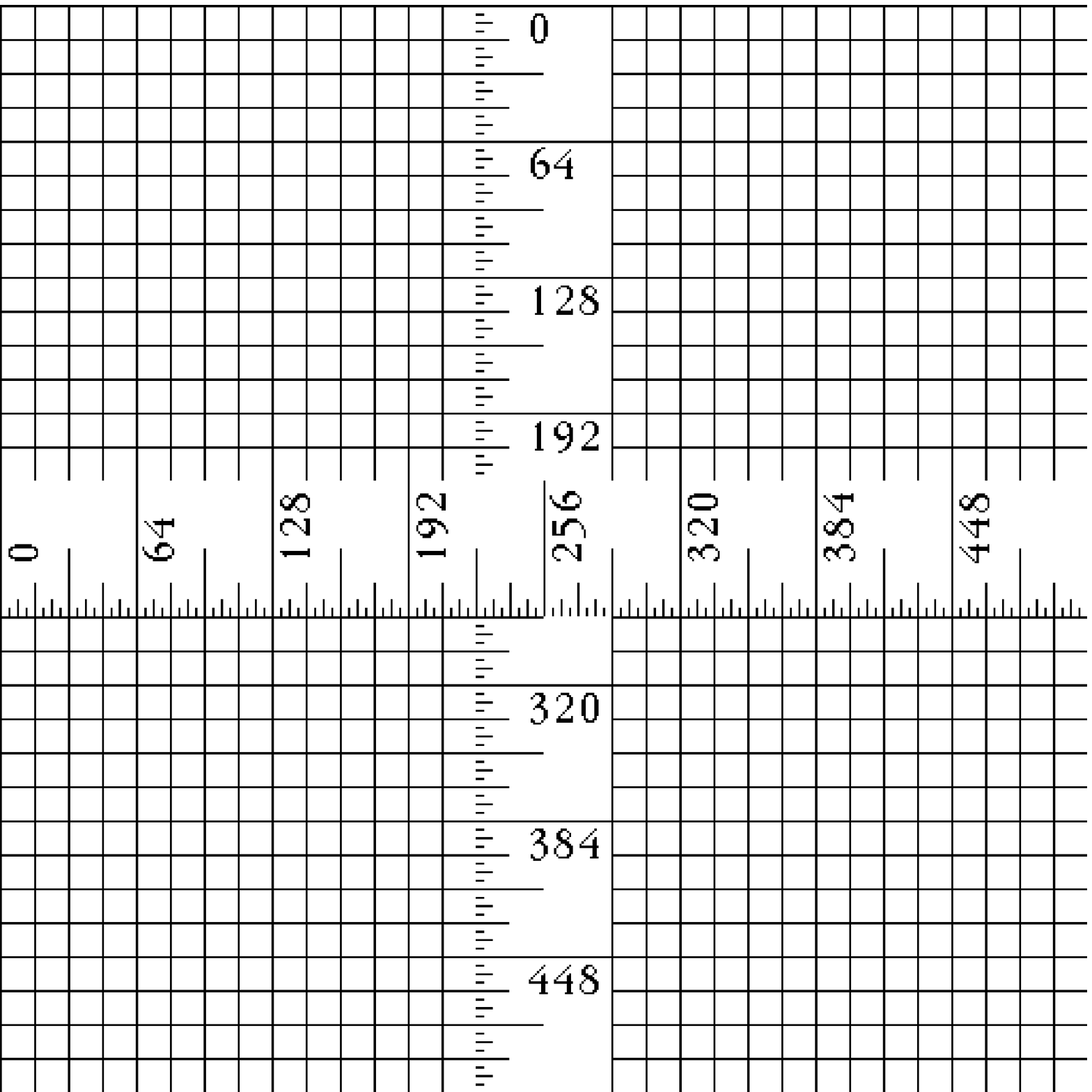, width=0.175\linewidth}}
\caption{Selected 512$\times$512 grayscale test images:
(a) Mandrill,
(b) Peppers,
(c) Couple,
(d) Aerial,
(e) Stream and bridge,
(f) Car and APCs,
(g) Fishing Boat,
(h) Girl (Elaine),
(i) Lena,
(j) USC texture mosaic \#1,
(k) Toy Vehicle, frame 1,
and
(l) Pixel ruler.}
\label{fig5}
\end{figure}

To compute the empirical pdfs of $\rho$,
an initial set of twelve standard images
available at
the USC SIPI database~\cite{USCSIPI} was selected,
as shown in Figure~\ref{fig5}.
Each image was
submitted
to the discussed watermarking procedure.
All the factors that contribute to the
quantification of the response variable $\rho$
were considered and, in a sense,
a factorial experiment was designed~\cite{hinkelmann2007experiments}.
Several wavelets were employed,
namely
Daubechies-4,
Daubechies-8,
Coiflet-6,
biorthogonal 6/2,
and
GRS4,
where the numbering
indicates the size of the respective filters.
Also various degrees of attack were utilized.
For the standard JPEG compression,
the attack strength can be measured by the
JPEG quality factor $Q$, which is inversely proportional
to the compression.
The selected quality factors were $Q\in\{10, 30, 50, 70, 90\}$.
Similarly,
for the JPEG2000 attack,
a range of different bit rates consisting
of 0.25, 0.5, 1, and 2 bits per pixels (bpp) was monitored.
After 300 trials for every image,
every wavelet,
every type of attack,
and every strength of attack,
the obtained values of $\rho$ constituted a
statistical sample
used
to derive the
sought empirical pdfs.

\begin{figure}
\centering
\subfigure[]{\epsfig{file=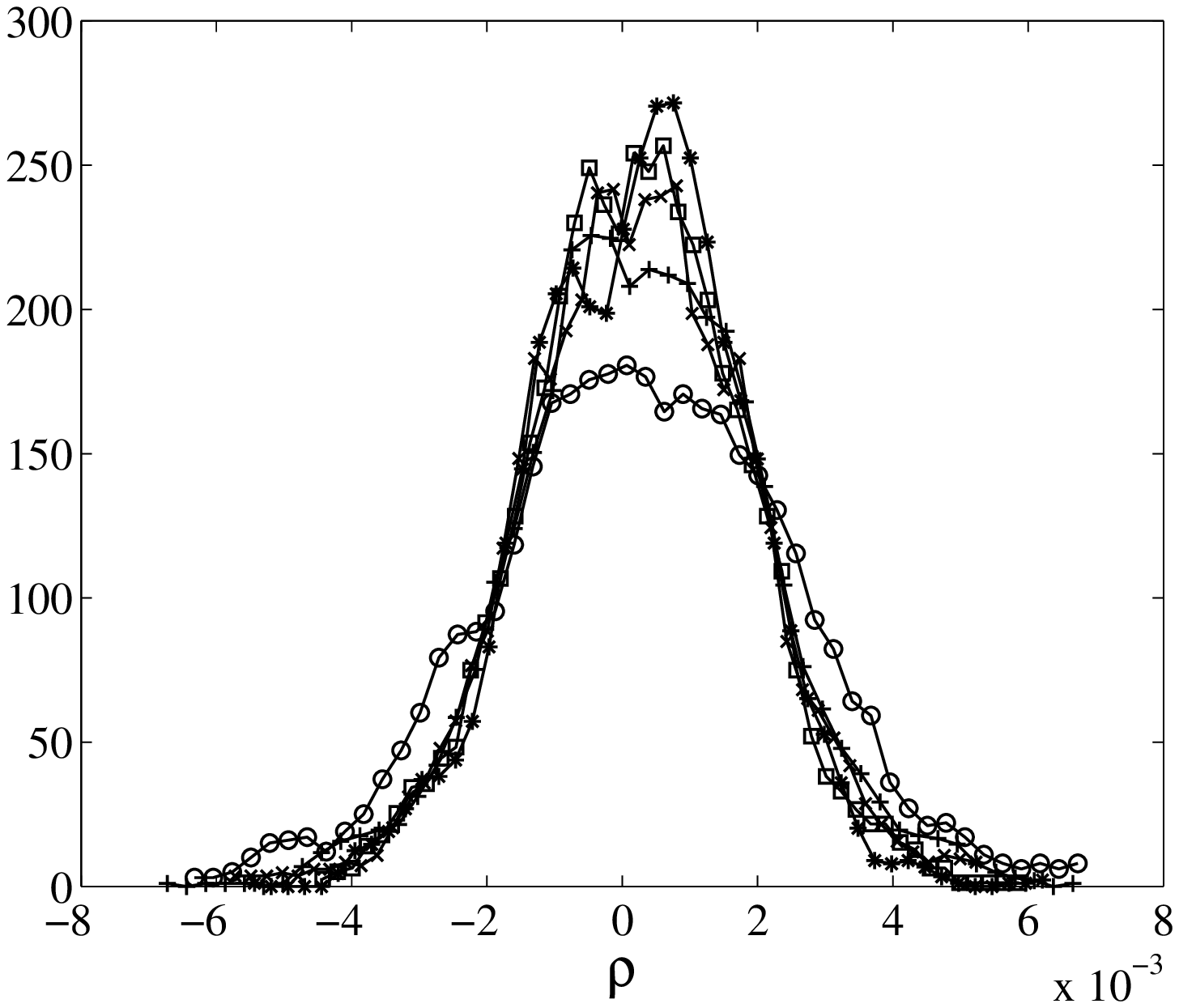, width=0.3\linewidth}}
\subfigure[]{\epsfig{file=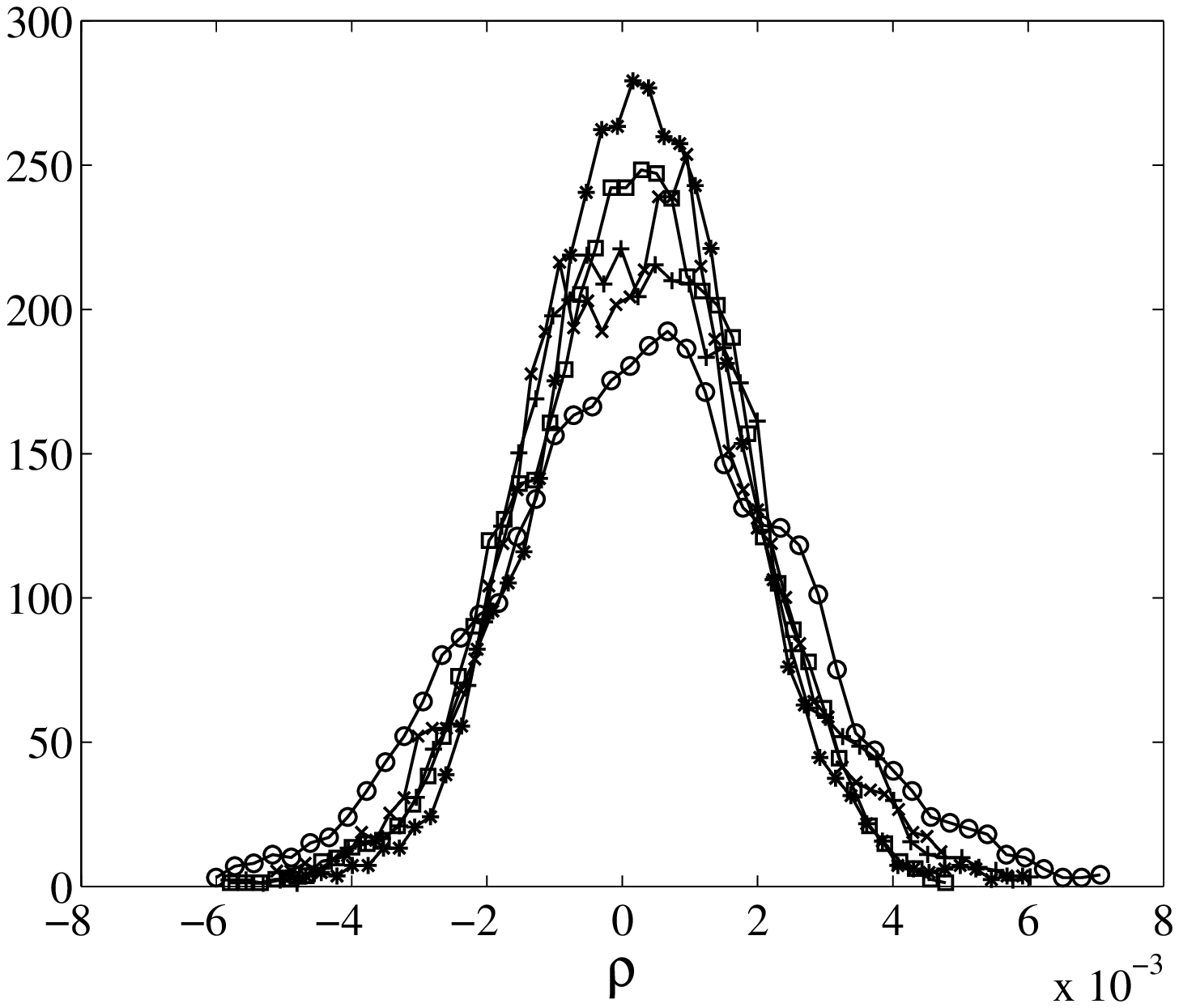, width=0.3\linewidth}}
\subfigure[]{\epsfig{file=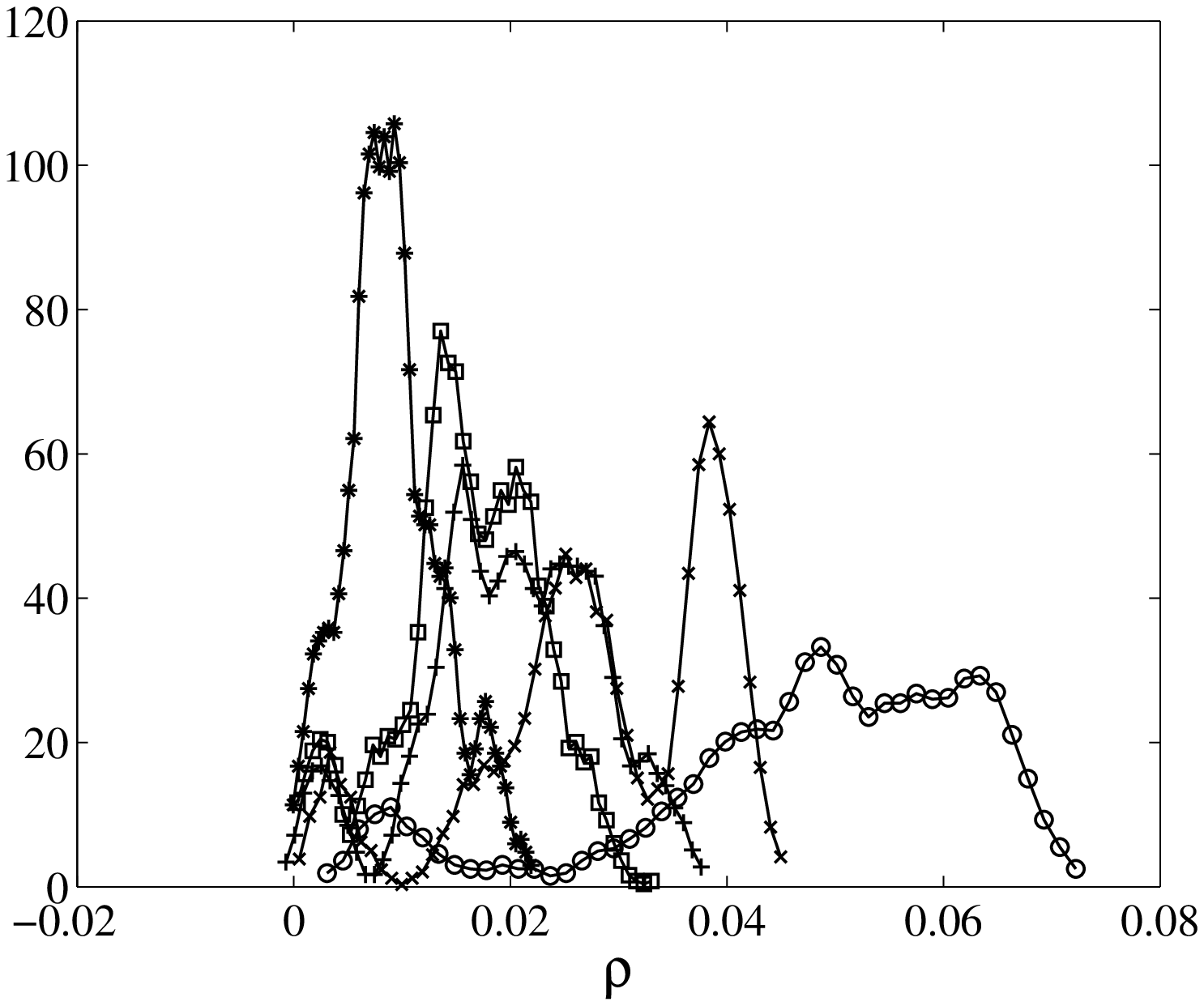, width=0.3\linewidth}}
\caption{Empirical pdfs of $\rho$ for the JPEG attack
with quality factor of 10\%
under
(a) null hypothesis with $\mathbf{W}=\mathbf{0}$,
(b) null hypothesis with $\mathbf{W}\neq\mathbf{0}$,
and
(c) alternative hypothesis.}
\label{fig-jpeg-10-epdf}
\end{figure}

\begin{figure}
\centering
\subfigure[]{\epsfig{file=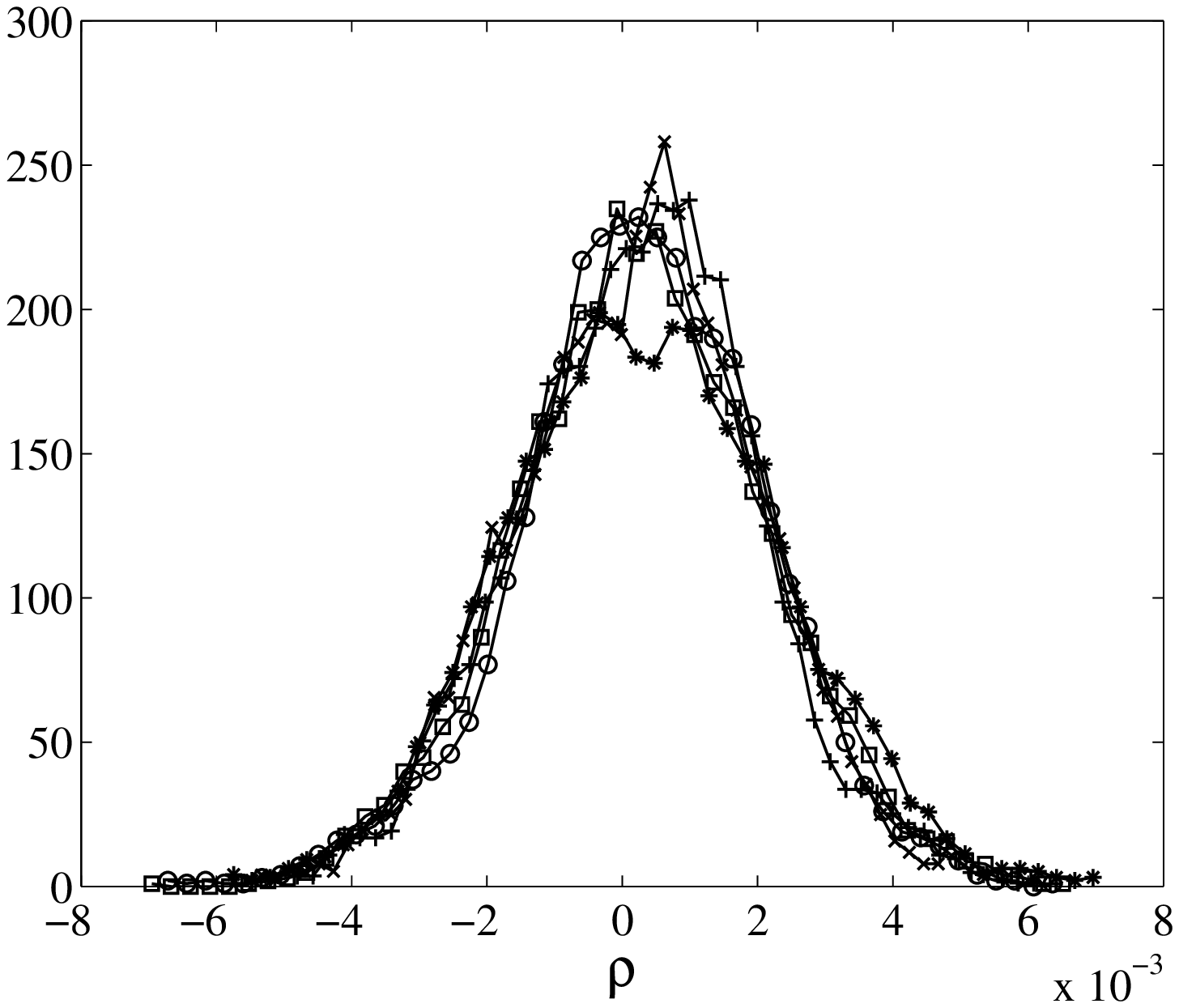, width=0.3\linewidth}}
\subfigure[]{\epsfig{file=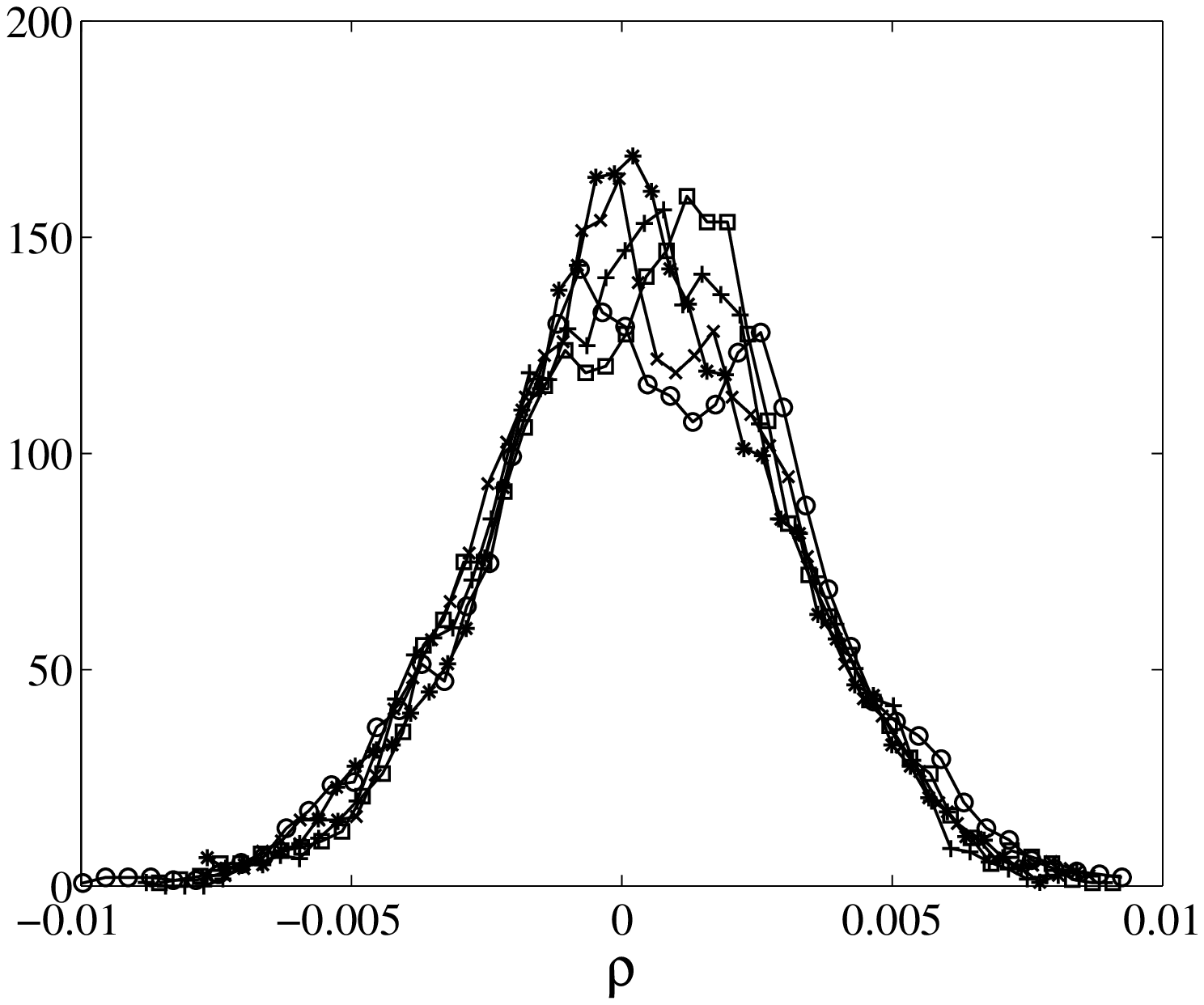, width=0.3\linewidth}}
\subfigure[]{\epsfig{file=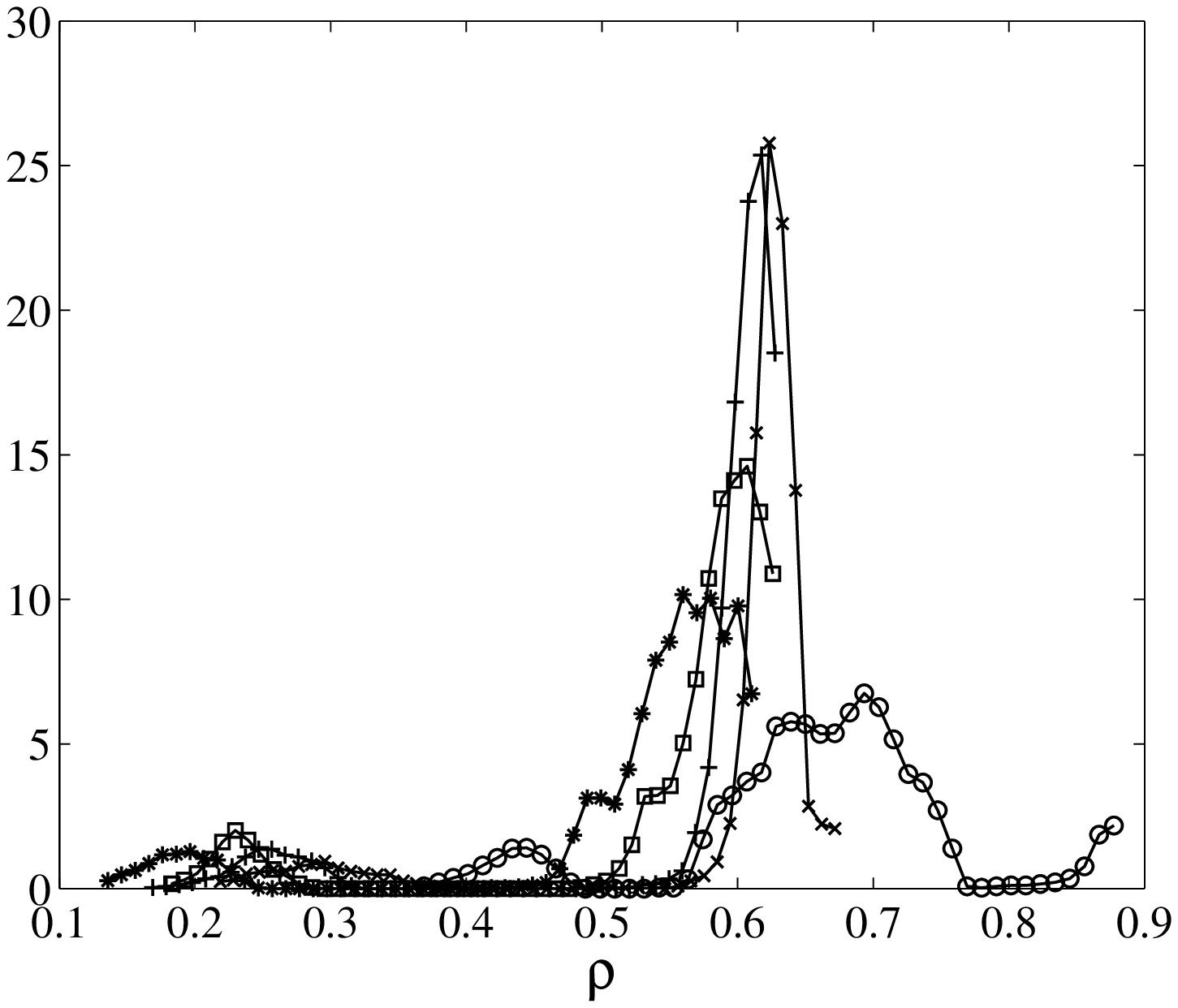, width=0.3\linewidth}}
\caption{Empirical pdfs of  $\rho$ for the JPEG attack
with quality factor of 90\%
under
(a) null hypothesis with $\mathbf{W}=\mathbf{0}$,
(b) null hypothesis with $\mathbf{W}\neq\mathbf{0}$,
and
(c) alternative hypothesis.}
\label{fig-jpeg-90-epdf}
\end{figure}

\begin{figure}
\centering
\subfigure[]{\epsfig{file=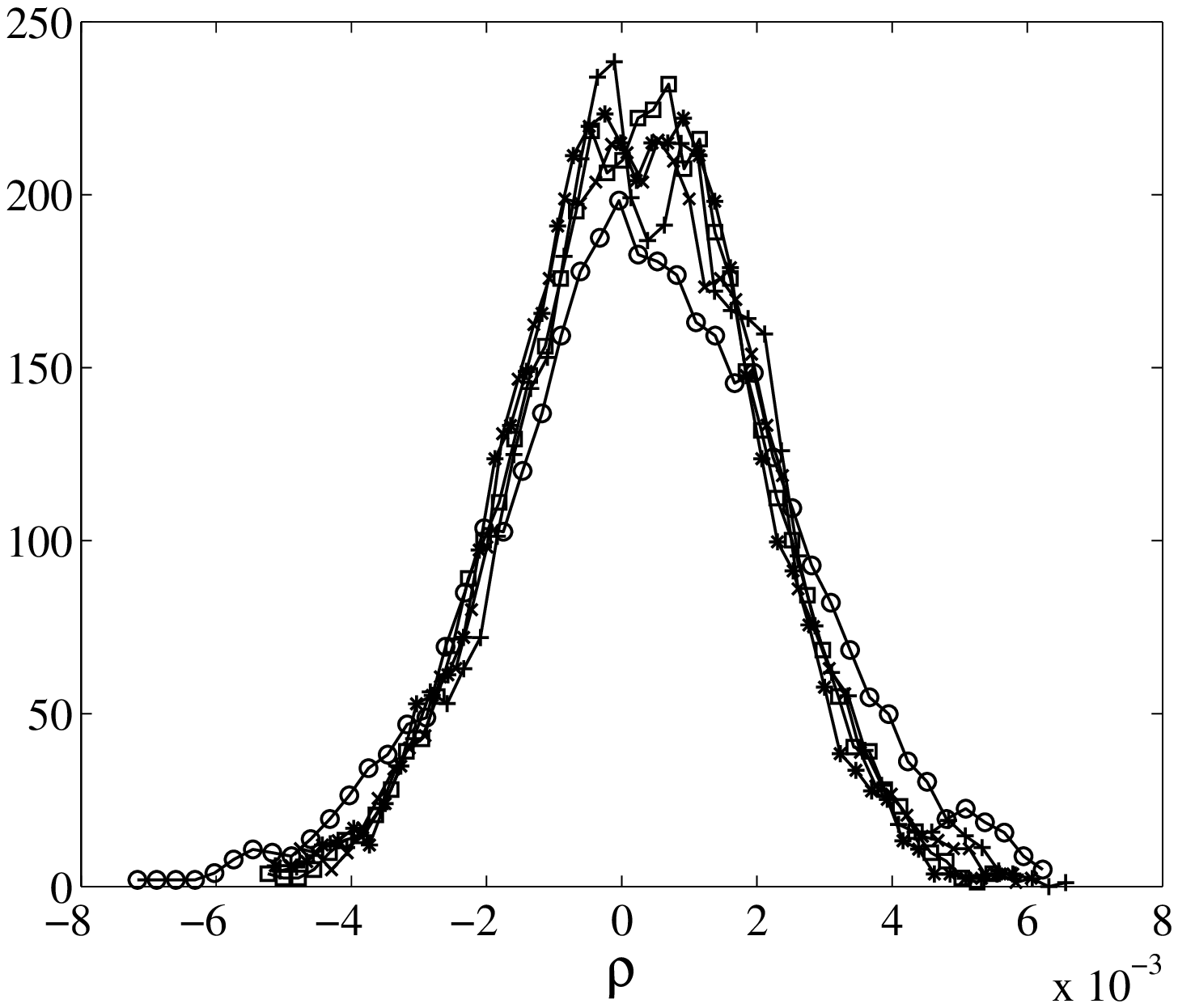, width=0.3\linewidth}}
\subfigure[]{\epsfig{file=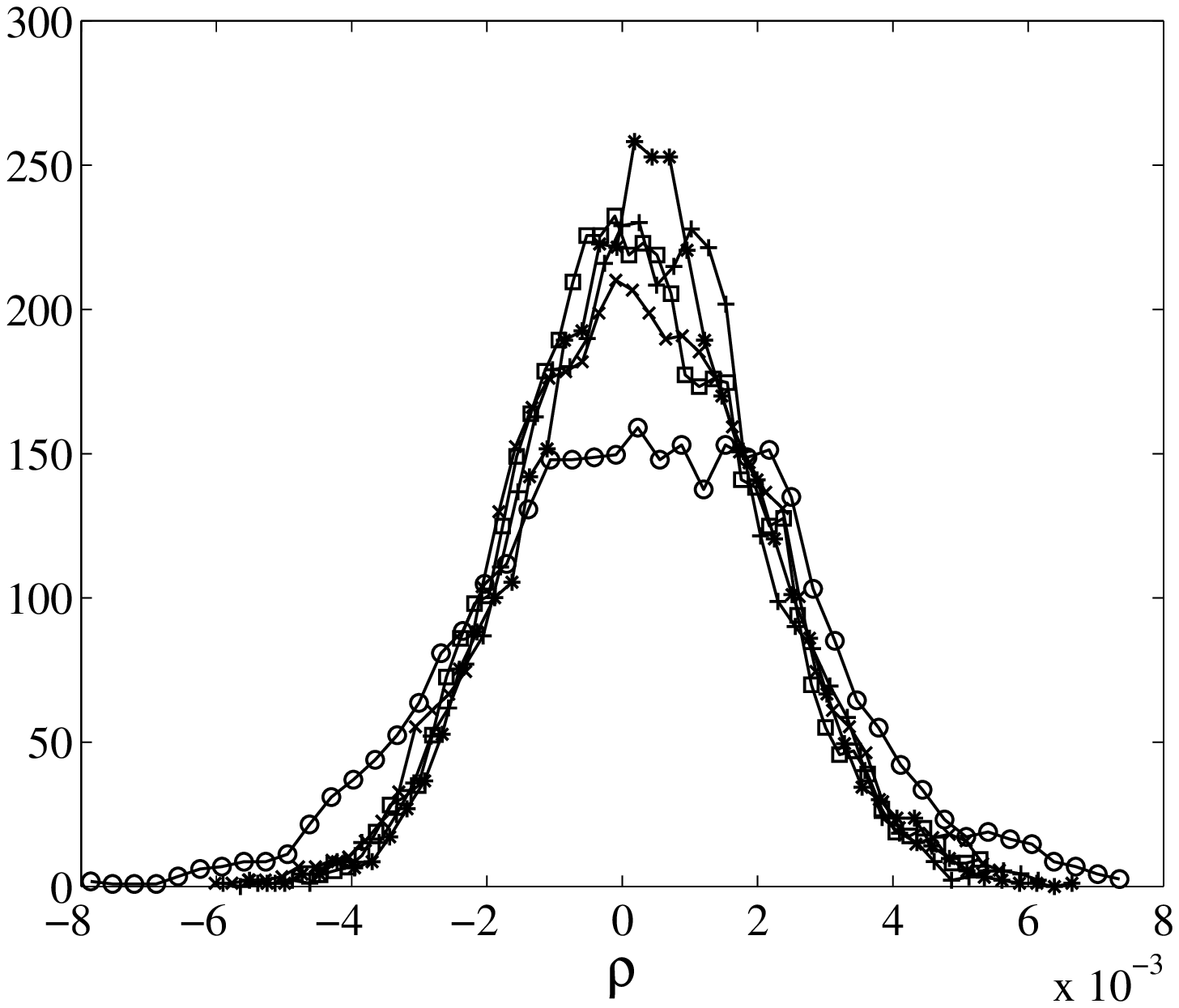, width=0.3\linewidth}}
\subfigure[]{\epsfig{file=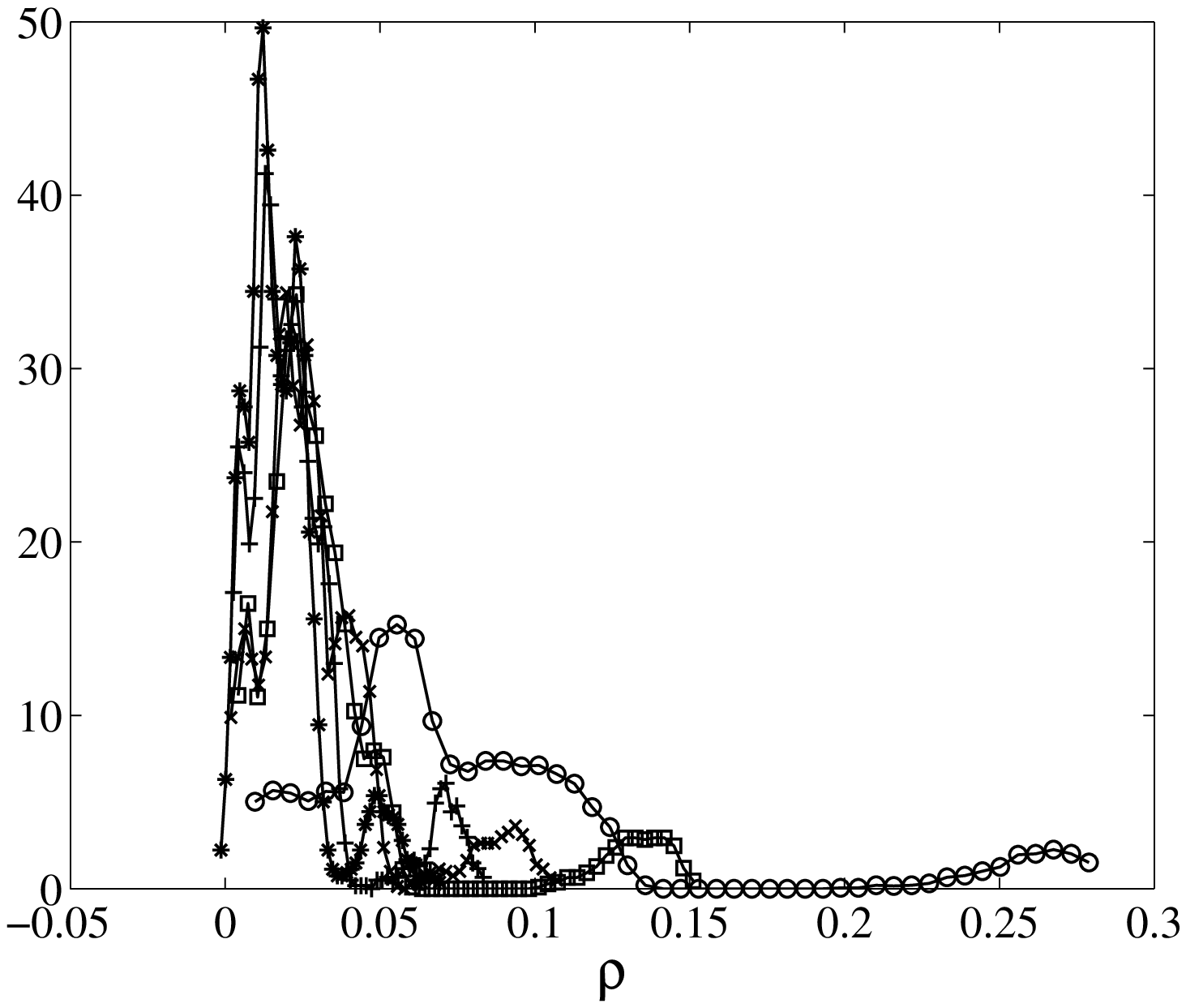, width=0.3\linewidth}}
\caption{Empirical pdfs of  $\rho$ for the JPEG2000 attack
at 0.25 bpp
under
(a) null hypothesis with $\mathbf{W}=\mathbf{0}$,
(b) null hypothesis with $\mathbf{W}\neq\mathbf{0}$,
and
(c) alternative hypothesis.}
\label{fig-jpeg2k-0.25bbp-epdf}
\end{figure}

\begin{figure}
\centering
\subfigure[]{\epsfig{file=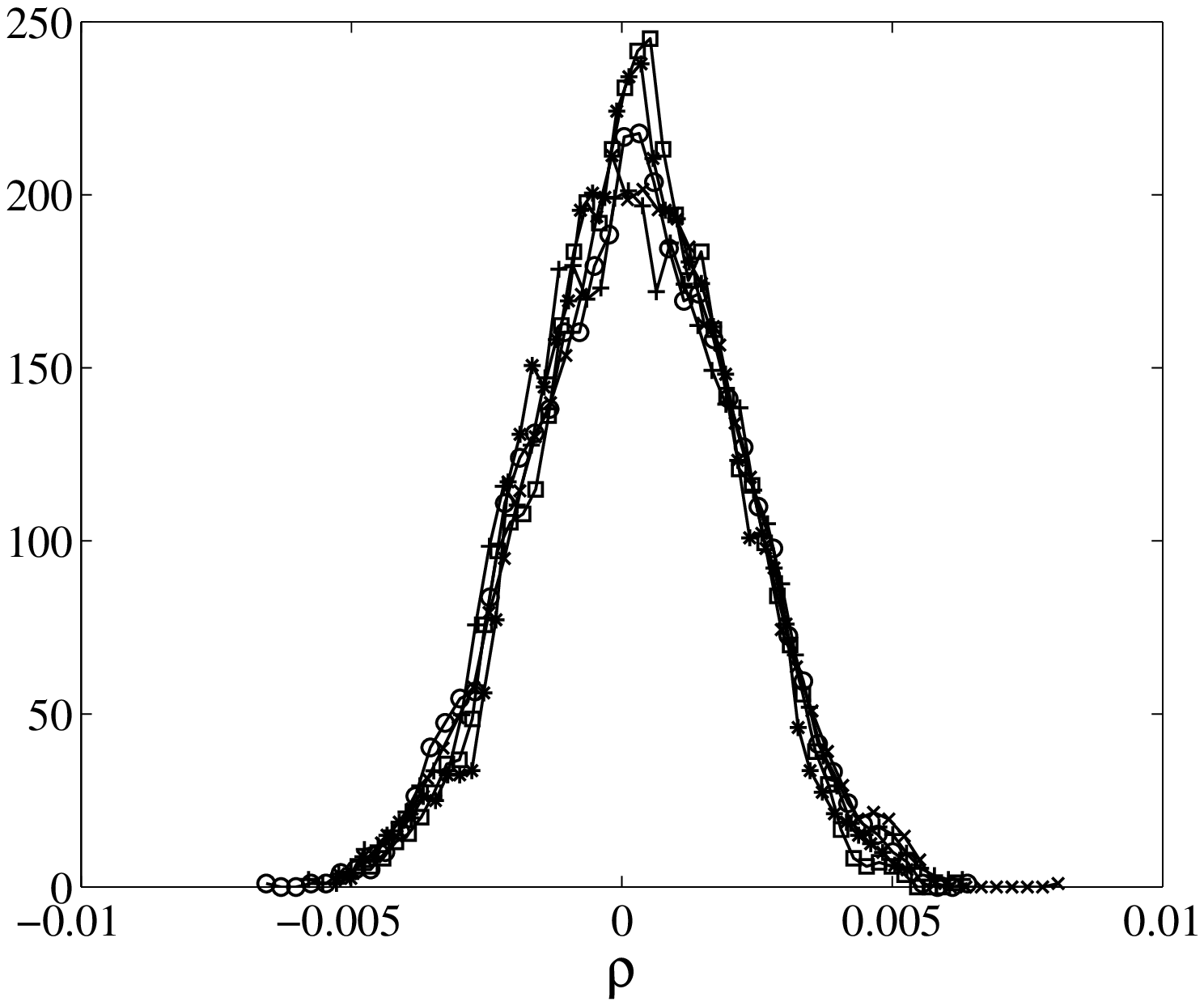, width=0.3\linewidth}}
\subfigure[]{\epsfig{file=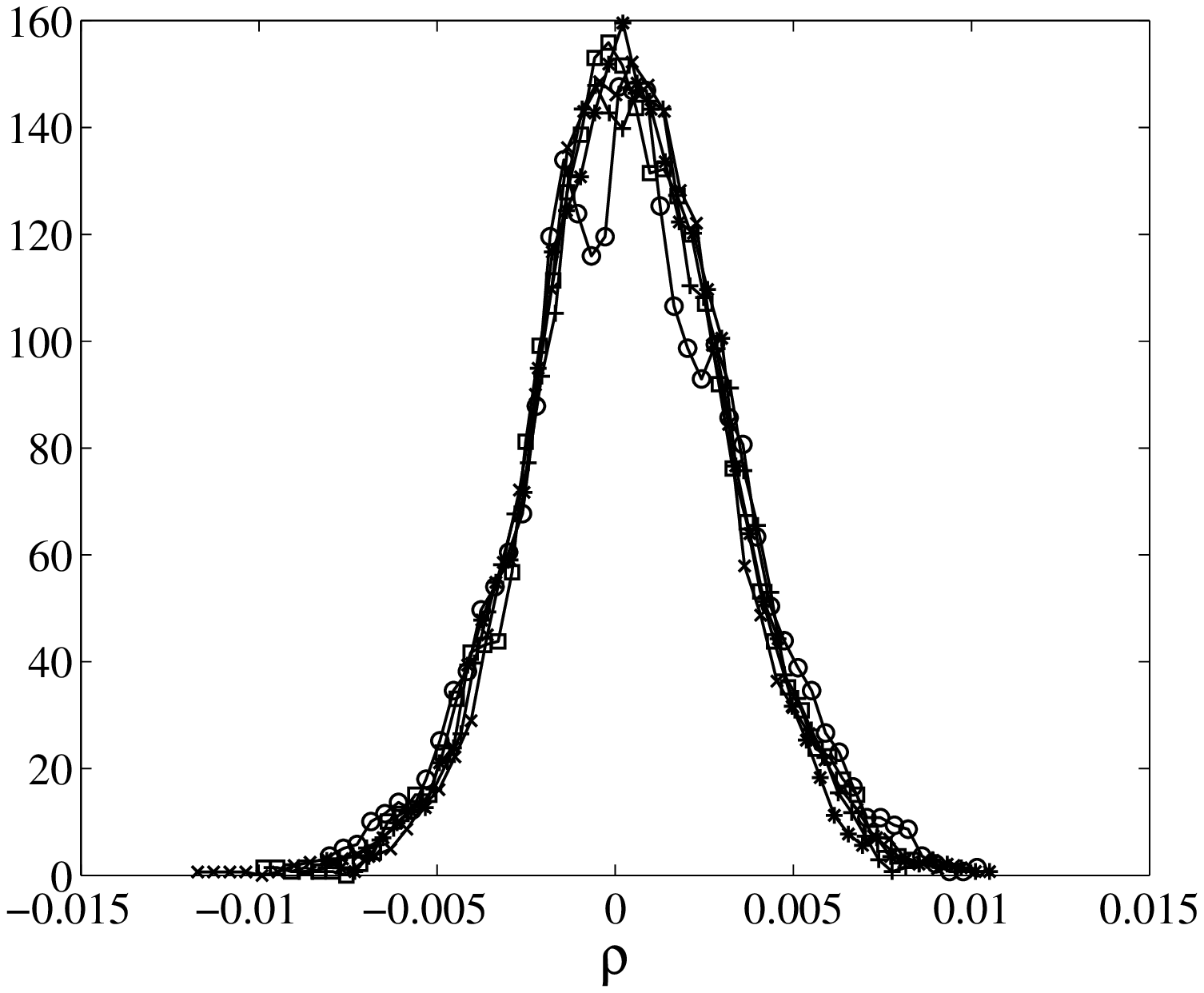, width=0.3\linewidth}}
\subfigure[]{\epsfig{file=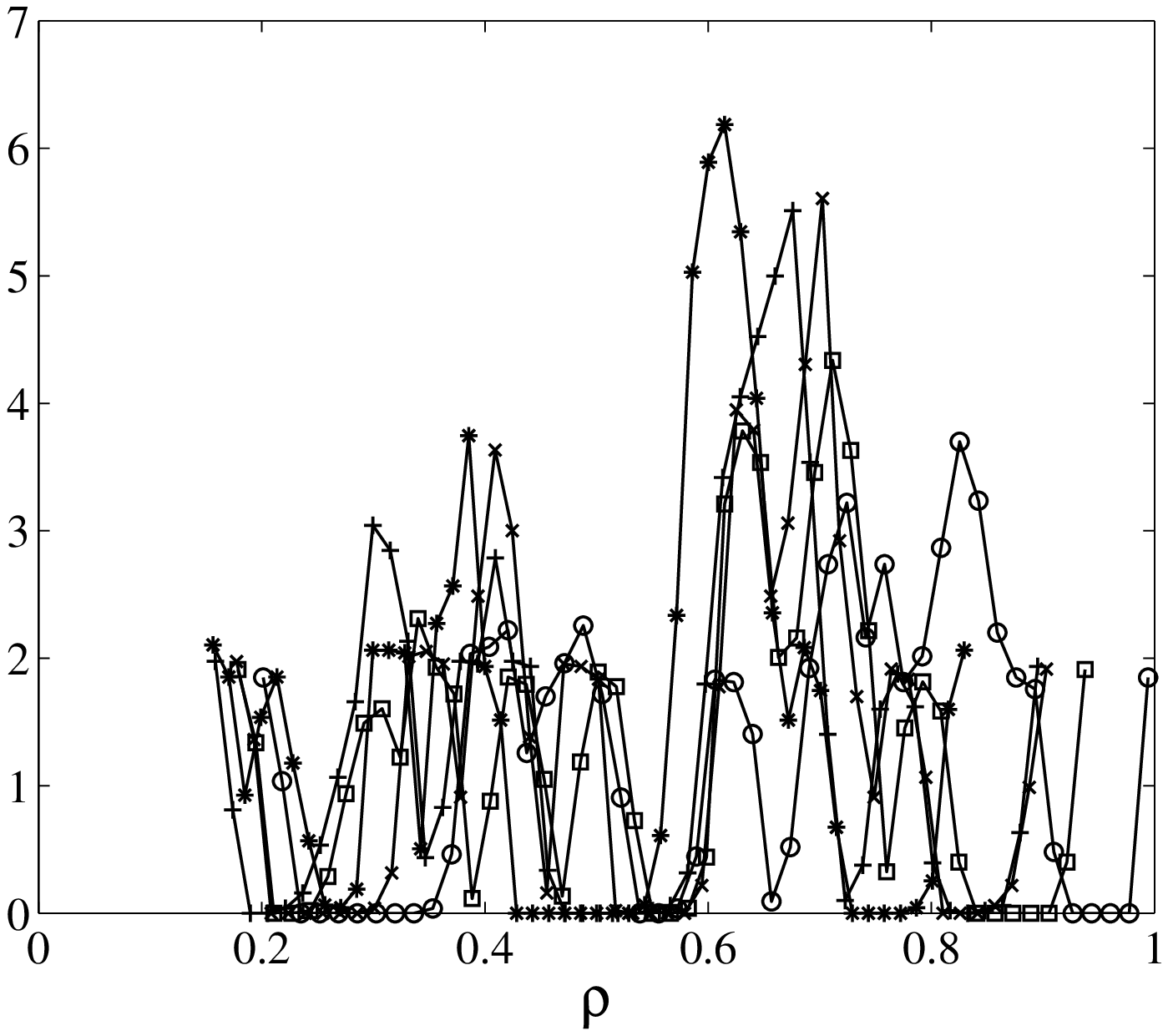, width=0.3\linewidth}}
\caption{Empirical pdfs of  $\rho$ for the JPEG2000 attack
at 2 bpp
under
(a) null hypothesis with $\mathbf{W}=\mathbf{0}$,
(b) null hypothesis with $\mathbf{W}\neq\mathbf{0}$,
and
(c) alternative hypothesis.}
\label{fig-jpeg2k-2bpp-epdf}
\end{figure}

The empirical pdfs for the JPEG attack at $Q=10$ and $Q=90$ are
respectively
displayed in Figures~\ref{fig-jpeg-10-epdf} and~\ref{fig-jpeg-90-epdf}.
Analogously,
Figures~\ref{fig-jpeg2k-0.25bbp-epdf} and~\ref{fig-jpeg2k-2bpp-epdf}
show the empirical pdfs for the JPEG2000 attack at
0.25~bpp and 2~bpp.
For conciseness, only the empirical pdfs associated to
the extremes cases were depicted.
Computational simulations
reveal
that the empirical pdfs for the intermediate degrees of attack
are, in qualitative terms,
proportionally situated in between the discussed
extreme cases.
Observe that, when an watermark mismatch occurred
(Figures~\ref{fig-jpeg-10-epdf}(a)-(b) and~\ref{fig-jpeg-90-epdf}(a)-(b);
and
Figures~\ref{fig-jpeg2k-0.25bbp-epdf}(a)-(b) and~\ref{fig-jpeg2k-2bpp-epdf}(a)-(b)),
the empirical pdfs possess a Gaussian-like shape,
indicating a manifestation of
the Central Limit Theorem.
The Lilliefors test for normality was applied and
Gaussianity hypothesis could not be rejected in all cases,
except in the following few situations:
(i) Daubechies-4 and biorthogonal 6/2 wavelets under JPEG attack at $Q=90$;
(ii) Coiflet-6 wavelet under JPEG attack at $Q=10$,
and
(iii) Daubechies-2 and GRS4 wavelets under JPEG2000 attack at 2 bpp.
Nevertheless, even for these cases,
the empirical pdfs resemble Gaussian pdfs,
indicating an eventual convergence to normality,
if sufficiently many trials were performed.

\begin{table}
\centering
\caption{Average and standard deviation of the empirical pdf of $\rho$
for several combinations of wavelets and JPEG attack strength}
\label{tab.jpeg}
\begin{tabular}{cccccc}
\hline
	& \multicolumn{5}{c}{Quality Factor}	 \\
\cline{2-6}
\raisebox{1.5ex}[0cm][0cm]{Wavelet} & 10 & 30 & 50 & 70 & 90  \\
\hline
Daubechies-4	& 0.0276$\pm$0.0108 & 0.0952$\pm$0.0272 & 0.1604$\pm$0.0503 & 0.2529$\pm$0.0666 & 0.5901$\pm$0.0956   \\
Daubechies-8	& 0.0193$\pm$0.0082 & 0.0746$\pm$0.0233 & 0.1322$\pm$0.0393 & 0.2342$\pm$0.0556 & 0.5714$\pm$0.0997   \\
Coiflet-6		& 0.0087$\pm$0.0045 & 0.0381$\pm$0.0155 & 0.0763$\pm$0.0271 & 0.1491$\pm$0.0432 & 0.5203$\pm$0.1095   \\
Biorthogonal 6/2	& 0.0154$\pm$0.0066 & 0.0588$\pm$0.0210 & 0.1095$\pm$0.0345 & 0.1993$\pm$0.0533 & 0.5531$\pm$0.1037   \\
GRS4		& 0.0458$\pm$0.0161 & 0.1395$\pm$0.0460 & 0.2289$\pm$0.0775 & 0.3379$\pm$0.1036 & 0.6537$\pm$0.0994   \\
\hline
\end{tabular}
\end{table}

\begin{table}
\centering
\caption{Average and standard deviation of the empirical pdf of $\rho$
for several combinations of wavelets and JPEG2000 attack strength}
\label{tab.jpeg2000}
\begin{tabular}{ccccc}
\hline
	& \multicolumn{4}{c}{Bit Rate (bpp)} \\
\cline{2-5}
\raisebox{1.5ex}[0cm][0cm]{Wavelet} 	& 0.25 & 0.5 & 1.0 & 2.0 \\
\hline
Daubechies-4	& 0.0275$\pm$0.0210 & 0.0817$\pm$0.0682 & 0.2212$\pm$0.1328 & 0.5559$\pm$0.2004\\
Daubechies-8	& 0.0207$\pm$0.0168 & 0.0637$\pm$0.0473 & 0.1920$\pm$0.1263 & 0.5261$\pm$0.2145\\
Coiflet-6		& 0.0167$\pm$0.0124 & 0.0537$\pm$0.0370 & 0.1649$\pm$0.0940 & 0.4870$\pm$0.2002\\
Biorthogonal 6/2	& 0.0311$\pm$0.0321 & 0.0902$\pm$0.0785 & 0.2384$\pm$0.1470 & 0.5574$\pm$0.2159\\
GRS4		& 0.0746$\pm$0.0618 & 0.1597$\pm$0.1240 & 0.3259$\pm$0.1918 & 0.6417$\pm$0.2251\\
\hline
\end{tabular}
\end{table}

Tables~\ref{tab.jpeg} and \ref{tab.jpeg2000} show ---
for the JPEG and JPEG2000 attack, respectively ---
the average and the standard deviation of
correlation coefficients.
Regardless the conditions,
the empirical pdfs related to the
non-regular wavelet GRS4 are
in the rightmost locations
as they exhibit greater averages.

The GRS4 filter offers the highest correlation coefficients in
all scenarios,
being followed by
the Daubechies-4 wavelet in the JPEG case
and by the biorthogonal 6/2 in the JPEG2000.
The GRS4 filter leads to the highest correlation coefficients
because watermarked messages using the GRS4 wavelet domain are more immune to compression
due the energy spreading property in the transform domain.
Effectively, the watermark is inserted simultaneously in low and high frequency subbands.
Therefore,
having in mind the most severe
attacks discussed above,
it is accepted that under less severe
attacks the detection performance
would be enhanced.
Then further analysis focuses
mainly the worst case scenario.

Subsequently,
the NP criterion was suitably adjusted to handle empirical pdfs
instead of analytically expressed pdfs.
Moreover,
numerical integration techniques were employed to solve
Equation~\ref{gamma} for the threshold~$\gamma$~\cite{hamming1986methods}.
Selecting a probability of false alarm $P_{FA} = 10^{-2}$,
the Neyman-Pearson approach
furnished the values of $\gamma$ shown in
Table~\ref{tab.gamma}.

\begin{table}
\centering
\caption{Threshold values for the detection algorithm}
\label{tab.gamma}
\begin{tabular}{ccc}
\hline
	& {JPEG @ $Q=10$}	& {JPEG2000 @ 0.25 bpp} \\
\hline
$\gamma$ & 0.340374 & 1.191534 \\
\hline
\end{tabular}
\end{table}

\section{Results and Discussion}
\label{section.results}

To assess the detection performance of the watermarking
scheme based on the GRS4 wavelet,
the probability of
detection and
the probability of
false alarm needed to be calculated.
By simulating the algorithm
a number of times,
the relative frequency of
detection $p_D$ and false alarm $p_{FA}$
can be used as estimators for
the sought probabilities.

For simulation purposes,
only the worst case scenario were taken in consideration,
namely JPEG compression for $Q=10$ and
JPEG2000 compression at 0.25 bpp.
Additionally,  we considered an entirely new set of eight images
from the USC SIPI database.
Otherwise, the use of previously examined images
would introduce bias to the results.
This set of images were submitted to
the proposed watermarking scheme with the GRS4 wavelet
for 8000 trials.
Table~\ref{tab.miss.ratio}
shows
the obtained relative frequencies.

\begin{table}
\centering
\caption{Detection and false alarm ratios}
\label{tab.miss.ratio}
\begin{tabular}{ccc}
\hline
Attack	& $p_D$ & $p_{FA}$ \\
\hline
JPEG @ $Q = 10$ & 0.9580  &  0.0089 \\
JPEG2000 @ 0.25 bpp & 1.0000 & 0.0020\\
\hline
\end{tabular}
\end{table}

We argue that the high frequency subbands of regular filters are less resistant
to compression because compression algorithms such as JPEG and JPEG2000 behave as lowpass filters.
As a result, watermark messages that are added
to the high frequency subbands are likely be filtered out by compression.
On the other hand, the subbands of non-regular transforms contain all frequency components.

To emphasize this point,
the correlation coefficient was computed separately
for each subband,
considering regular and non-regular wavelets.
Again examining the worst case scenario,
we compared, in the JPEG attack at $Q=10$,
the results derived from
the Daubechies-4 and the GRS4 wavelets.
Analogously,
for the JPEG2000 attack at 0.25 bpp,
the biorthogonal wavelet was
set against the GRS4 wavelet.
These particular wavelets were elected
because they granted the largest values of
average $\rho$ in their respective
empirical pdfs,
as shown in Tables~\ref{tab.jpeg} and~\ref{tab.jpeg2000}.

\begin{figure}
\centering
\subfigure[]{\epsfig{file=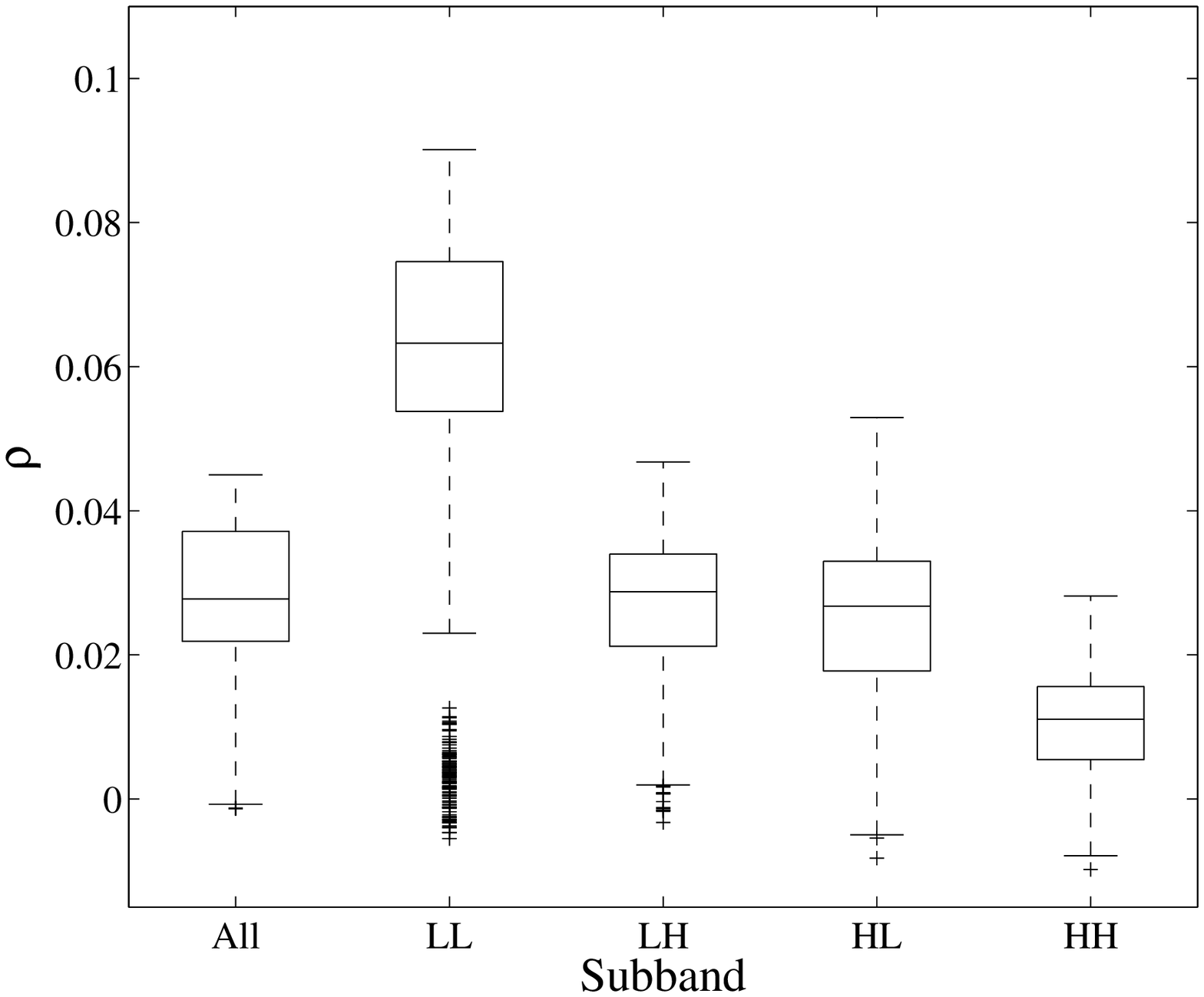, width=0.4\linewidth}}
\subfigure[]{\epsfig{file=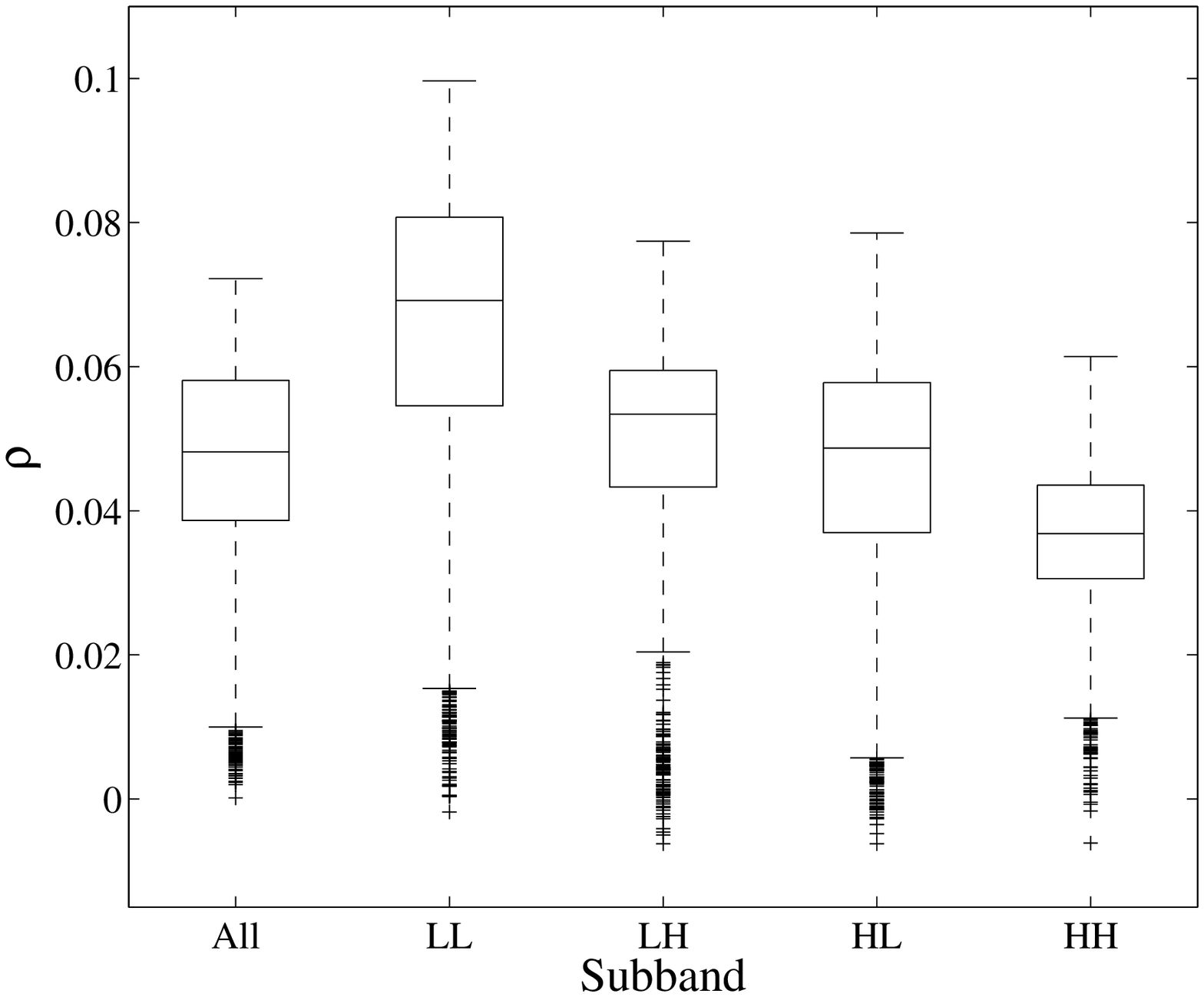, width=0.4\linewidth}}
\caption{Box and whisker plot of $\rho$ for JPEG attack at $Q=10$ categorized by subbands.}
\label{boxplot.jpeg}
\end{figure}

\begin{figure}
\centering
\subfigure[]{\epsfig{file=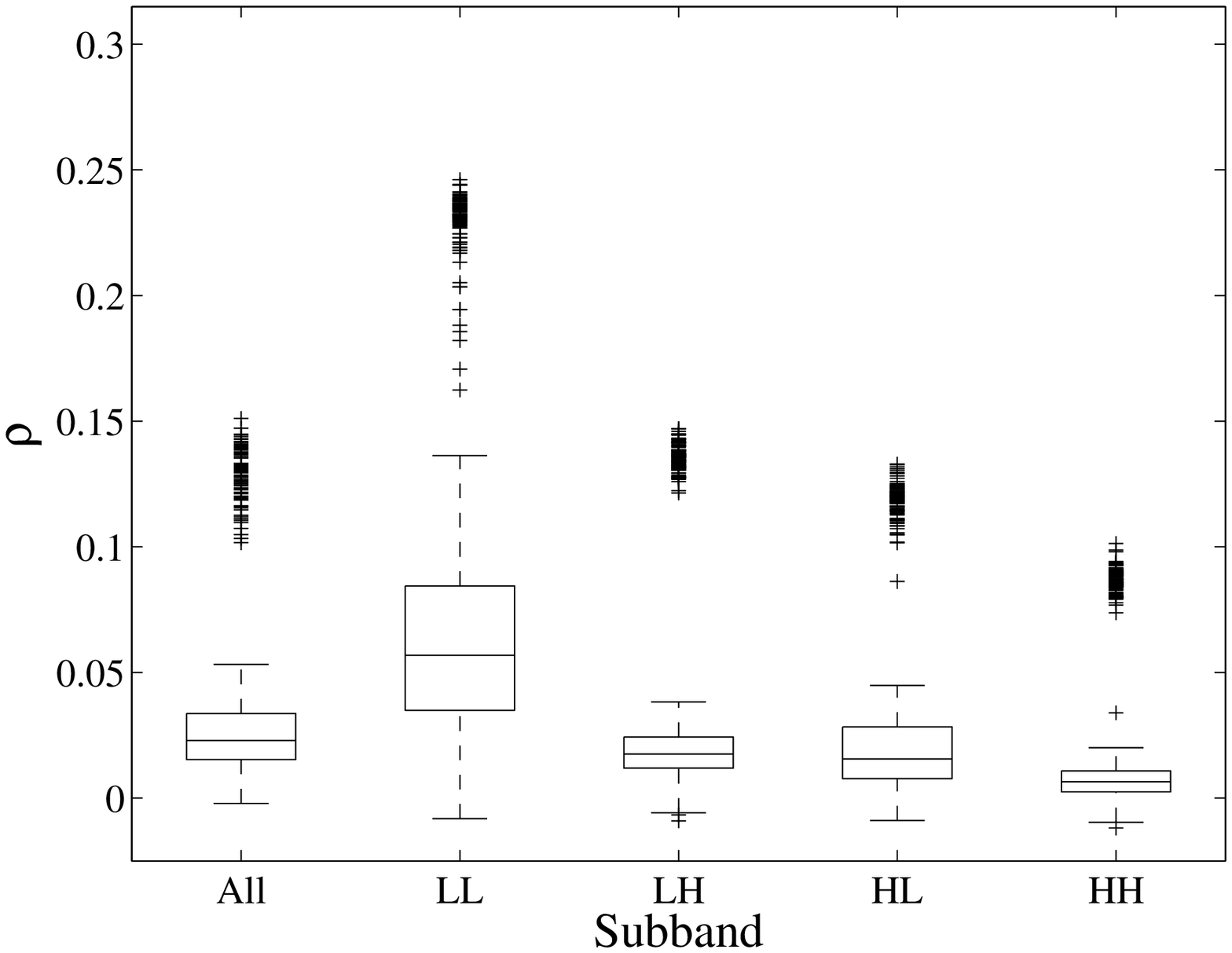, width=0.4\linewidth}}
\subfigure[]{\epsfig{file=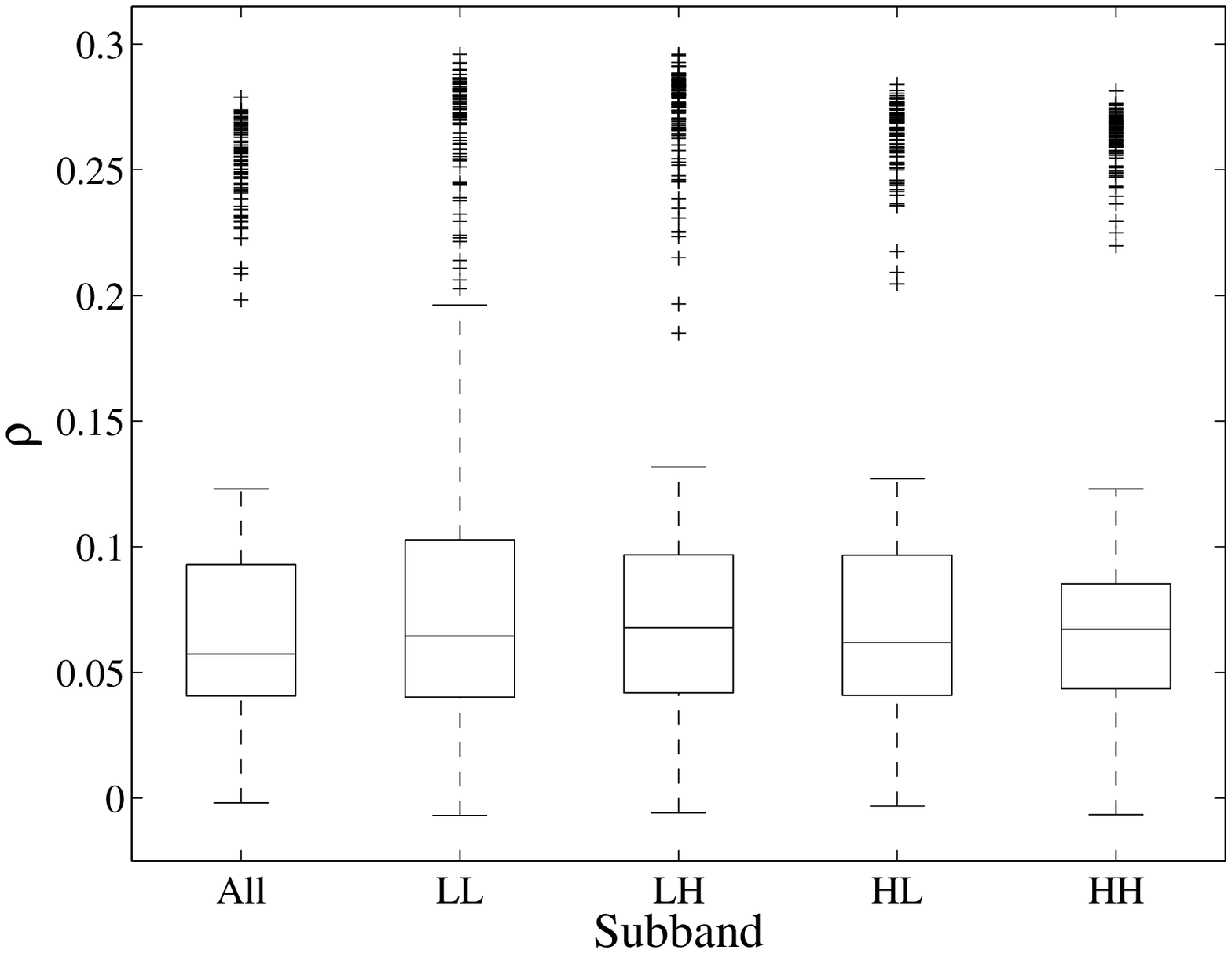, width=0.4\linewidth}}
\caption{Box and whisker plot of $\rho$ for JPEG2000 attack at 0.25 bpp categorized by subbands.}
\label{boxplot.jpeg2000}
\end{figure}

As a result,
Figures~\ref{boxplot.jpeg} and~\ref{boxplot.jpeg2000}
depict box and whisker plots of $\rho$
categorized by subband.
Note that in all subbands
the average value of $\rho$ for the
non-regular wavelets is greater than
the quantities offered by the regular wavelets.
The correlation coefficients of the LL subbands
are generally
superior when compared with other subbands.
This result is not surprising as most of the energy is
concentrated in the LL subband.
More than that,
when non-regular wavelets are utilized,
the variance of the average $\rho$ throughout
the subbands are considerably lower when compared
with the regular wavelets.
This indicates that
the information stored in the higher subbands is
significant for watermarking detection when
non-regular wavelet are employed.
Computational evidence also demonstrates that the above discussed
behavior remains valid for the other values of
quality factor and bit rate considered in this work.

\section{Conclusions}
\label{section.conclusions}

In this paper non-regular wavelet filters are shown to have useful properties for image watermarking.
These filters provide energy spreading.
As a result, at the same time image quality is high and the watermark is
very robust to lossy operations furnished by the JPEG and JPEG2000 compression schemes.

{\small
\bibliographystyle{IEEEtran}
\bibliography{nonregular}
}

\end{document}